\pdfoutput=1
\documentclass[aps,prb,showpacs,reprint,numarical,superscriptaddress]{revtex4-1}
\usepackage{float}
\usepackage{natbib}
\usepackage{amsmath}
\usepackage{graphicx,xcolor,xhfill,calc,xspace}
\usepackage{amssymb}
\usepackage{tikz,pgfplots}
\usepackage{standalone}
\usetikzlibrary{plotmarks,calc,shapes, positioning}
\usepgflibrary{arrows}
\usepackage[mathscr]{euscript}
\usepackage{hyperref}
\usepackage{cleveref}
\definecolor{sqfill}{rgb}{0.77333,0,0}
\definecolor{sqborder}{rgb}{0.97333,0,0}
\definecolor{cfill}{rgb}{0.622,0.453,0.253}
\definecolor{cborder}{rgb}{0.466,0.346,0.2} 

\definecolor{orange}{rgb}{0.7,0.2,0}
\definecolor{darkgreen}{rgb}{0,0.5,0}
\definecolor{grey}{rgb}{.5,0,0.4}

 \DeclareRobustCommand\circ{\begin{tikzpicture} \fill [blue,draw = blue!80, scale = 0.8] (0ex,0ex) circle (1ex); \draw [blue!80] (-3ex,0) -- (-1ex,0); \draw [blue!80] (1ex,0) -- (3ex,0); \end{tikzpicture}}
 \DeclareRobustCommand\square{\begin{tikzpicture} \fill [sqfill, draw = sqborder,scale = 0.8] (-1ex,-1ex) rectangle (1ex,1ex); \draw [sqborder] (-3ex,0) -- (-1ex,0); \draw [sqborder] (1ex,0) -- (3ex,0);  \end{tikzpicture}}
\DeclareRobustCommand\ccirc{\begin{tikzpicture} \fill [cfill,draw = cborder,scale = 0.8] (0ex,0ex) circle (1ex); \draw [cborder] (-3ex,0ex) -- (-1ex,0ex); \draw [cborder] (1ex,0) -- (3ex,0ex);  \draw [cborder] (-0.707ex,-0.707ex) -- (0.707ex,0.707ex); \draw [cborder] (-0.707ex,0.707ex) -- (0.707ex,-0.707ex);
 \end{tikzpicture}}

\begin{document}
\title{Guided magnonic Michelson interferometer}
\author{ Muhammad H.~Ahmed}
\affiliation{Chemical and Quantum Physics, School of Applied Sciences, RMIT University, Melbourne 3001, Australia}

\author{Jan Jeske} 
\affiliation{Chemical and Quantum Physics, School of Applied Sciences, RMIT University, Melbourne 3001, Australia}

\author{ Andrew D.~Greentree}
\affiliation{Chemical and Quantum Physics, School of Applied Sciences, RMIT University, Melbourne 3001, Australia}
\affiliation{ARC Centre of Excellence in Nanoscale BioPhotonics, School of Science, RMIT University, Melbourne, VIC 3001, Australia}
 \date{\today}

%\affil[*]{hamiid9@gmail.com}

%\keywords{Keyword1, Keyword2, Keyword3}

\begin{abstract}
Magnonics is an emerging field with potential applications in classical and quantum information processing. Freely propagating magnons in two-dimensional media are subject to dispersion, which limits their effective range and utility as information carriers. We show the design of a confining magnonic waveguide created by two surface current carrying wires placed above a spin-sheet, which can be used as a primitive for reconfigurable magnonic circuitry. We theoretically demonstrate the ability of such guides to counter the transverse dispersion of the magnon in a spin-sheet, thus extending the range of the magnon. A design of a magnonic directional coupler and controllable Michelson interferometer is shown, demonstrating its utility for information processing tasks.
\end{abstract}
\maketitle

\section{Introduction}

 During the past two decades, magnonics has become a major field of theoretical and experimental research \cite{Kruglyak:2010cy,Lenk:2011el,Chumak:2015fa}. The potential use of spin waves in several technologies, like computing (quantum \cite{Khitun:2001bva} and classical \cite{Chumak:2014iw,Khitun:cb,Khitun:2010dx}), communication \cite{Vogt:2014he} and caloritronics \cite{Bauer:2012fq} has been investigated thoroughly. Magnon-magnon scattering also provides an interesting test bed for exploring  particle-particle interactions more generally \cite{Bose:2007ji,Longo:2013kk}. The core idea of magnonics is to manipulate spin waves in ferromagnetic or antiferromagnetic materials, and in doing so to produce devices with different functionalities \cite{Lenk:2011el,Chumak:2014iw,Schneider:2008fu}. This manipulation is usually achieved by the application of external fields \cite{Serga:2009dg,Chumak:2009fz,Neumann:2009hz}, by crafting the material itself (removing the material on sides the achieve a physical channel) \cite{Chumak:2009eg,Anonymous:f5yEVrch,Chumak:2009koa,Vasiliev:2007br} or some combination of both \cite{Schneider:2008fu,Kostylev:2005fua}.

     In a magnonic device there is no net transport of charge as opposed to electronic devices, where current flow leads to inevitable heating.  For this reason, magnonic technology promises the advantage of reduced power consumption compared to its electronic counterparts. \cite{Chumak:2015fa,Khitun:2010dx}

     A spin wave is a propagating magnetic disturbance in an otherwise perfectly ordered magnetic medium \cite{Dyson:1956kb,Heisenberg:1928jw}. When quantised, the associated particle is a magnon. Here we are concerned with the propagation of magnons confined to a two dimensional array of spin-1/2 particles (spin sheet). Within the sheet, each spin interacts with its nearest neighbour through exchange or dipole-dipole coupling. Due to this interaction they either align (ferromagnetic coupling) or anti-align (anti-ferromagnetic coupling) themselves to their nearest neighbours, resulting in a global order in spin direction as the lowest energy state. Any local disturbance in the global order creates propagating magnons in the sheet\cite{Kruglyak:2010cy}.
     
    Freely propagating magnons are subject to damping and dispersion, which limits the distances over which the magnon can propagate \cite{Serga:2010cw,vanHoogdalem:2011bo,Jeske:2013cs}. The effects of dispersion can be mitigated by confining the magnon in the direction transverse to the travel direction. Typically, such confinement is achieved by lithographically removing material to leave a magnonic channel  \cite{Chumak:2009koa,Vasiliev:2007br,Chumak:2009eg,Anonymous:f5yEVrch}, which minimises the transversal dispersion. However, such approaches are not suited to reconfigurable devices due to the permanent structural changes induced in the material.

  \begin {figure}[hb!]
    \resizebox{\columnwidth}{!}{
    \includegraphics[trim  = 0mm 10mm 10mm 0mm, clip ]{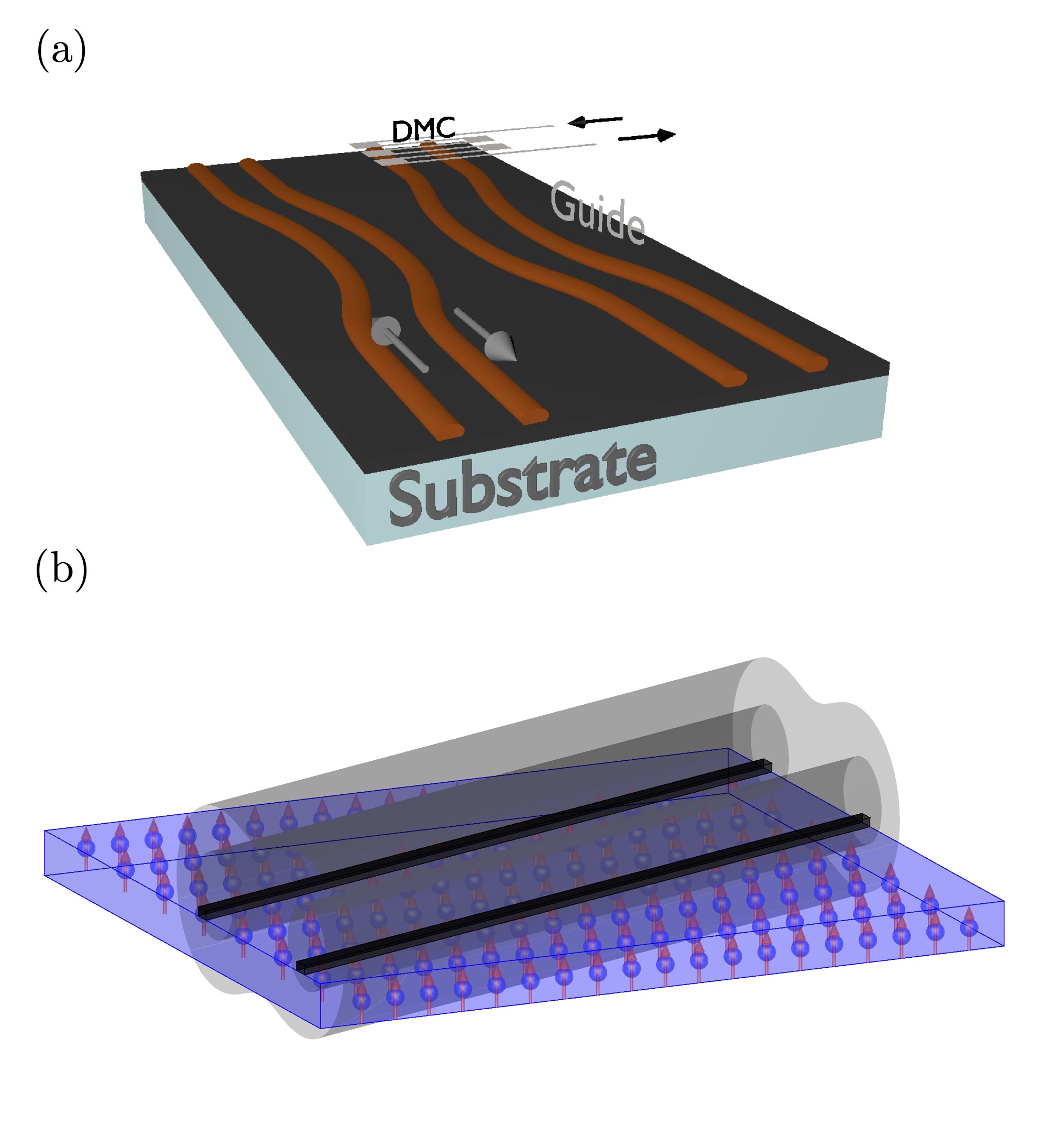}
    }
    \caption{(a) Schematic of spin guide. Black lines represent current carrying wires with current flowing in opposite direction. Black translucent surfaces are magnitude field contours. The surface wires define a confining potential in the direction of the wire. (b) Schematic of the magnonic Michelson interferometer. The black layer shows a two-dimensional ferromagnetic sheet on a substrate. Brown lines are current carrying wires and the arrows show the direction of the current in the wires. One arm has a dynamic magnonic crystal (DMC) at its end for the purpose of phase control. }
    \label{F1}
    \end{figure}

    Here we describe a method of confining and transporting a magnon in a two-dimensional ferromagnetic film, which does not require any structural modification of the film itself. In our model, the confinement is achieved by the temporally invariant magnetic field from two parallel current-carrying wires placed on the surface at some distance from each other. Schematics of such a device is shown in \cref{F1} (a), where \cref{F1} (b) shows the design of magnonic Michelson Interferometer.

     When the current direction in the wires is opposite, the resulting magnetic field forms a potential well, creating a channel for the magnons, which is termed a spin-guide. Previously we have shown analogous confinement in one-dimensional spin chains using a surface-gate array, where the transport was achieved through adiabatic temporal variation of the potential \cite{Makin:2012vi,Ahmed:2015eo}. 

    %Our results are directly relevant to the design of magnonic circuits, especially those which utilise coherent properties. 
   
    Our results enable reconfigurable magnonic circuitry of several, potentially interacting, magnon channels that preserve  phase and thereby allow quantum information processing. The channels can be reconfigured by adapting the current in the wires that create the channel. One can consider multiple spin-guides on a single surface that can be switched on or off as desired. This ability can be used to build complex magnonic circuitry for all-magnon information processing. More generally, our work is motivated by efforts to demonstrate traditional quantum optical effects in non-traditional tight-binding systems\cite{Makin:2012vi,Nikolopoulos:2008ka,Yung:2005gm,Ahmed:2015eo,Makin:2009ja}.
    
    To illustrate the utility of our scheme, we present a full design for a guided magnon version of a Michelson interferometer.  The Michelson interferometer is an important element for high-precision sensing and can also be used as a switching primitive for all-magnon logic gates \cite{Klingler:2014hk}.
    
    This paper is set out as follows. We first introduce the Heisenberg Hamiltonian in two dimensions and a magnon confining potential. We then consider the extension to a directional coupler and finally a magnonic Michelson Interferometer.

\section{Free and guided magnon propagation in a 2D Heisenberg sheet}
   \begin{figure*}[ht!]
 \centering
  \resizebox{1.8\columnwidth}{!}{
  \includegraphics{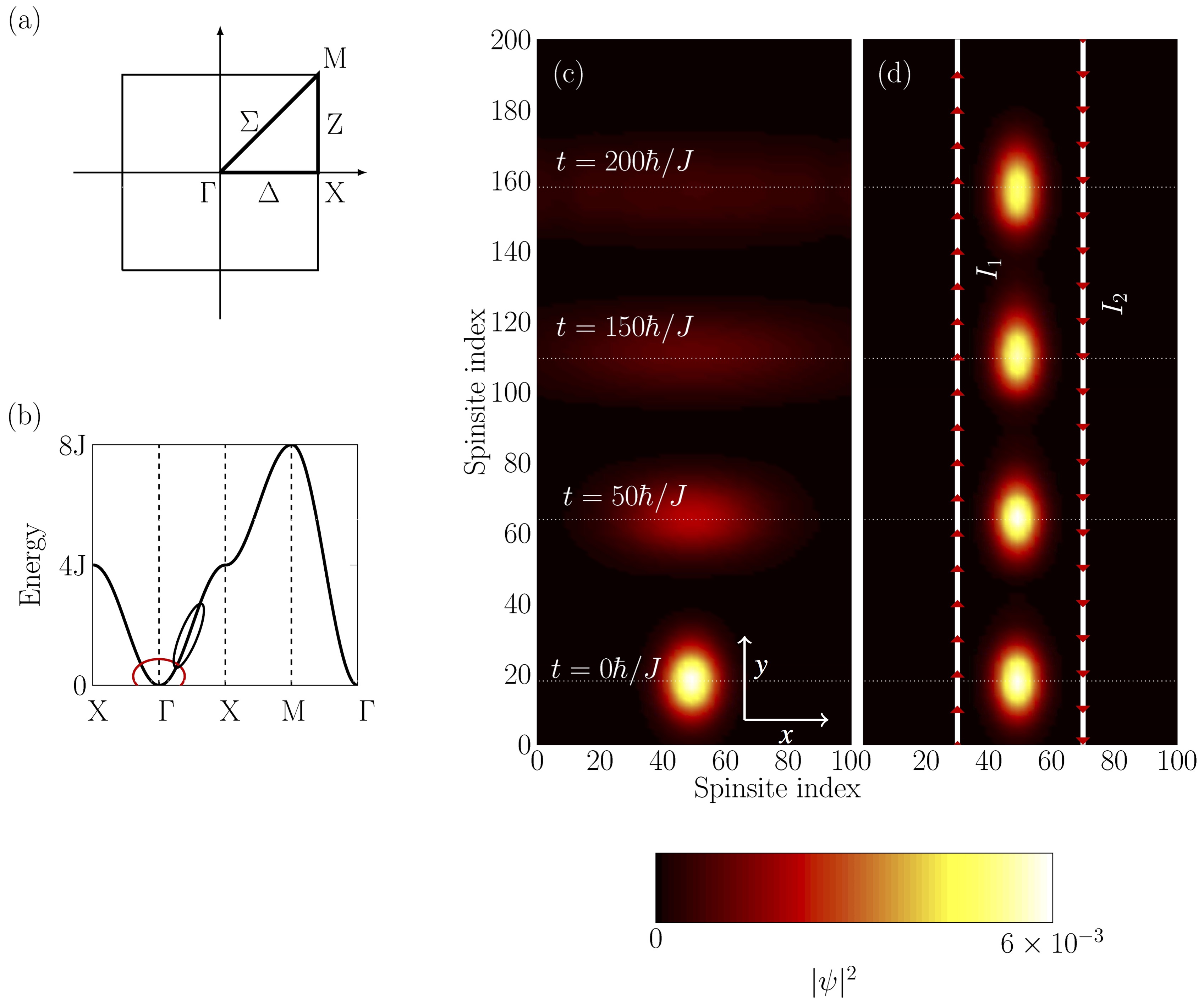}
};

     \caption {(a) Brillouin zone of a square lattice with the lines of high symmetry marked in bold. (b) Dispersion relation of the square lattice along high symmetry lines. Red and black ellipses show the areas of high and low dispersion, respectively. (c) Overlaid snap shots of a propagating magnon in a square-lattice, while the color axis is $|\psi|^2$. Rapid transversal dispersion is apparent where the longitudinal dispersion is not noticeable at this distance. This is because the magnon was chosen with a $y$-momentum, which is in a low dispersion region, as marked in black in (b). (d) Magnon confined by the surface current-carrying wires for the same initial state as (c). Red arrows show the direction of the current in the wires and white lines represent wires (d = 20$a$, $w_g = 20a \ \textrm{and} \ \varepsilon_{\textrm{min}}/J = 0.1$). After the confinement there is no transverse spreading of the magnon.} 
\label{disp_n_prop}
 \end {figure*}

    To calculate the magnon propagation, we introduce the Heisenberg Hamiltonian with nearest neighbour interaction in a two dimensional square lattice, with $\hbar =1$. 
    \begin{align}
    \label{hamiltonian} 
    H=-&J \sum_{i,j}\big[{\mathbf{S}}_{i,j}.{\mathbf{S}}_{i+1,j} +{\mathbf{S}}_{i,j}.{\mathbf{S}}_{i,j+1} \big]+  \sum_{i,j} \varepsilon_{(i,j)},\ \     
    \end{align} 
    
\noindent $\mathbf{S} = (\sigma^x, \sigma^y, \sigma^z)$ is the total spin operator, where $\sigma^x, \sigma^y \textrm{and} \ \sigma^z$ are the Pauli spin operators. $J$ is the strength of nearest neighbour interaction. The nature of the interaction can be exchange or dipole-dipole. If both types of couplings are present then $J$ becomes the effective coupling with the magnitude equal to the sum of both coupling. i.e.~ $J = J_{ex} + J_d$, $J_{ex}$ is the exchange coupling and $J_d$ is the dipole-dipole coupling\cite{deSousa:2004dua}.

The third term in the Hamiltonian [\cref{hamiltonian}] is the on-site energy of $(i,j)$th spin due to the local external magnetic field $B_{i,j}$, with $\varepsilon_{i,j} = \gamma B_{i,j}\sigma^z_{i,j}$, where $\gamma$ is the gyromagnetic ratio.

 We do not take into account the second-nearest neighbours i.e. $(i,j) \ \textrm{to}\  (i+1,j+1)$ coupling. For magnons propagating close to parallel to either of the principle axes, the effects of such couplings can be approximated by an effective increase of the nearest-neighbour coupling. Through an analytical derivation (see Appendix) we demonstrate that the effect of these interactions is equivalent to that of having a stronger $J$ coupling, for magnons travelling in either $x$ or $y$-direction. For such a case, to account for second-nearest-neighbour coupling, we substitute $J = J_e+2J_d$ in \cref{hamiltonian}, where $J_e$ is the standard edge coupling and $J_d$ is the diagonal coupling on a 2D square lattice.
   
 The dispersion relation of a freely propagating magnon in a two dimensional sheet is  
    \begin{align}
\omega_{k_{x},k_{y}} =  &= 4J - 2J[ \textrm{cos}(k_{x}a)+\textrm{cos}(k_ya)],
    \label{dispersion}
    \end{align}  
  % \boldsymbol  
    \noindent where $\omega_{k_{x},k_{y}}$ is the angular frequency of a particular eigenmode, $k_x$ and $k_y$ are the components of $\boldsymbol{K} $, wave numbers along $x$ and $y$ directions and $a$ is the spin-spacing. The dispersion curve of a square lattice along the high symmetry lines is shown in \cref{disp_n_prop}(a,b). 
    
 To illustrate the free magnon dispersion we considered a magnon with the initial wave function

\begin{align} 
    \psi(x_i,y_j) = & \exp\left[-\frac{(x_0-x_i)^2}{2\phi_x^2}-\frac{\left(y_0-y_j\right)^2}{2\phi_y^2}\right]   \nonumber \\
     & \exp({-ik_xx_i-ik_yy_j}). \label{packet}
 \end{align}

\noindent \Cref{packet} corresponds to a two-dimensional Gaussian wavepacket propagating in the $x$ and $y$-direction with the initial group velocity corresponding to wave vector $\boldsymbol{K}  =  k_x \hat{x} + k_y \hat{y}$. $\phi_x$ and $\phi_y$ are the spatial standard deviations in the $x$ and $y$-directions and ($x_0$, $y_0$) is the center of the wave packet. 
 
 In an experimental setup, the standard way to excite magnons in a thin film is by the microwave induction technique \cite{Chumak:2015fab,Chumak:2014iw,Demidov:2011jo,Gieniusz:2013eb}. A surface antenna is subjected to AC current and the Oersterd field of the antenna introduces propagating spinwaves inside the sheet. The frequency and the wavelength of the excitation can be controlled directly by the frequency of the AC current. When such an antenna is used on the top of the guide to excite magnons, only the confined modes will travel along the guide and all unconfined modes will leak out and disperse away. This approach can be thought of as being analogous to coupling into a few-mode optical fiber.

    %In the case of free magnon propagation, choice of gaussian wave-packet is completely arbitrary and any other shape will follow similar kind of dynamics.
    
    The required $\boldsymbol{K}$-vector to achieve a certain group velocity of the magnon can be calculated by the following relation 
       \begin{align}
    %    \label{gvel}
    %    \omega &= \frac{2J}{\hbar}\left[1-\cos((k_x+k_y)a)\right], \\
        \label{vg}
         \boldsymbol{V_{g}} = &  v_{g} \hat{x}+v_{g} \hat{y}  =\partial_{k_x} \omega_{k_{x},k_{y}} \hat{x} \ +  \partial_{k_y} \omega_{k_{x},k_{y}}\hat{y} \nonumber \\
          = & 2Ja[\sin(k_xa) \hat{x} + \sin(k_ya) \hat{y}].  
        \end{align}    
In our simulation we initiated the magnon with a y-velocity $\boldsymbol{V_g} = 2aJ \hat{y}$, which corresponds to $k_ya = \pi/2$ and $k_x=0$.  
  
 \Cref{disp_n_prop}(c) shows the evolution of a free magnon along the spin sheet. The time evolution was achieved through solving the discrete version of the Schr{\"o}dinger equation and hard wall boundary conditions were implemented. We observe strong dispersion in the transverse direction and weak dispersion in the lateral direction. This is due to the fact that the two directions correspond to the different parts of the dispersion curve. In the propagation direction, the group velocity is closer to the linear part [\cref{disp_n_prop}(b) black ellipse] of the dispersion curve, hence it shows reduced dispersion, whereas the zero momentum region is a high dispersion region [\cref{disp_n_prop}(b) red ellipse] and hence spreading is more rapid.

  \begin{figure*}[t!]
  \centering
  \resizebox{1.8\columnwidth}{!}{
  \includegraphics{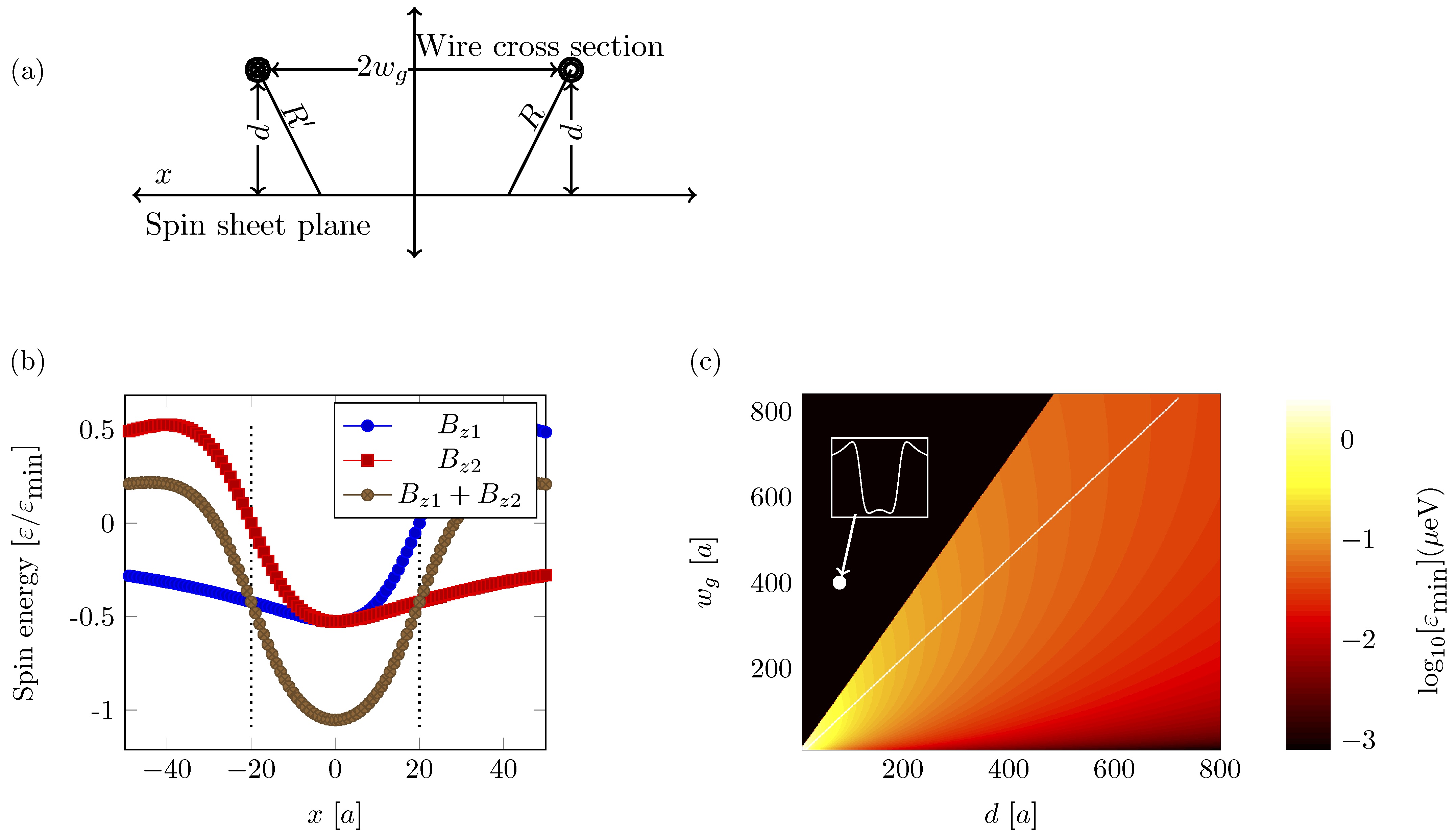}
};
\caption{(a) Cross section of a spin-guide. Two circles show the position of the wires, where the inner circle and cross inside each circle signifies the direction of the current. The horizontal line is the plane of the spin sheet. $2w_g$ is the separation between wires and $d$ is the distance between the spin sheet and the wires. (b) Onsite energy of spins inside the sheet due to individual (\ \square \  and \ \circ\ ) and combined (\ \ccirc \ ) magnetic field of two current carrying wires. Two vertical (dotted) lines show the position of the wires. The two wires form a potential well with depth $\varepsilon_{\textrm{min}}$, which can guide magnons.(c) A pseudocolor plot of the depth, $\varepsilon_{\textrm{min}}$, due to the magnetic field of the wires, located at the center of the potential. The black region is the region where the potential starts to split into two separate potentials and therefore leads to more complicated dynamics. The inset shows the shape of such a split potential with the parameters that lie inside the black region($d = 80\textrm{a}, w_g = 400\textrm{a}$). The white line marks the position of the maximum potential depth.}
  \label{shape_n_depth}
\end{figure*}

 For our confinement scheme, we use two current carrying wires with equal but antiparallel currents, placed at some distance above the plane of the spin-sheet, as shown in \cref {shape_n_depth}(a). \Cref{disp_n_prop} (d) shows the evolution of a magnon inside the guide. As expected, there is negligible transverse dispersion, however the longitudinal dispersion is unchanged. The white lines show the wires and the red arrows show the direction of current in the wires. The longitudinal initial state for the confined case were same as of the free case. The transversal profile was chosen to be the ground state of the confining potential. 

In addition to the field of the wires, we also considered a global z-field, which is large compared to the wires' magnetic field. This large field energetically separates the total ground state from the single-excitation subspace, and also allows the secular approximation.  The secular approximation allows us to neglect all terms from non-$z$ components of the wire-field because they are averaged out by the strong $z$-field. Hence we perform all calculations in the single excitation subspace and ignore the $x$ and $y$-magnetic field components due to the wires.
The functional form of the combined z-field of the wires at some point $x$ inside the sheet is  
    
   \begin{align}
   B_z(x) = \frac{\mu_0 I}{2\pi} \Bigg[ \frac{x-w_g}{R^2}-\frac{x+w_g}{R'^2}\Bigg] \label{f_form}, \ 
     \end{align}  
      
\noindent where $R =  \sqrt{d^2+(x-w_g)^2}$,  $R' =  \sqrt{d^2+(x+w_g)^2}$, $d$ is the distance between the sheet plane and wires, $w_g$ is the half-width of the guide (half-separation between the wires), $\mu_0$ is the permeability of free space and $I$ is the current in the wires. \Cref{shape_n_depth}(a) shows a cross section of the sheet and wires. Two circles represent the wires running perpendicular to the plane of the page and the horizontal line is the plane of the spin-sheet. The current direction in each wire is opposite to each other. 
  \begin{figure*}[t]
 \centering
 \resizebox{1.6\columnwidth}{!}{
   \includegraphics{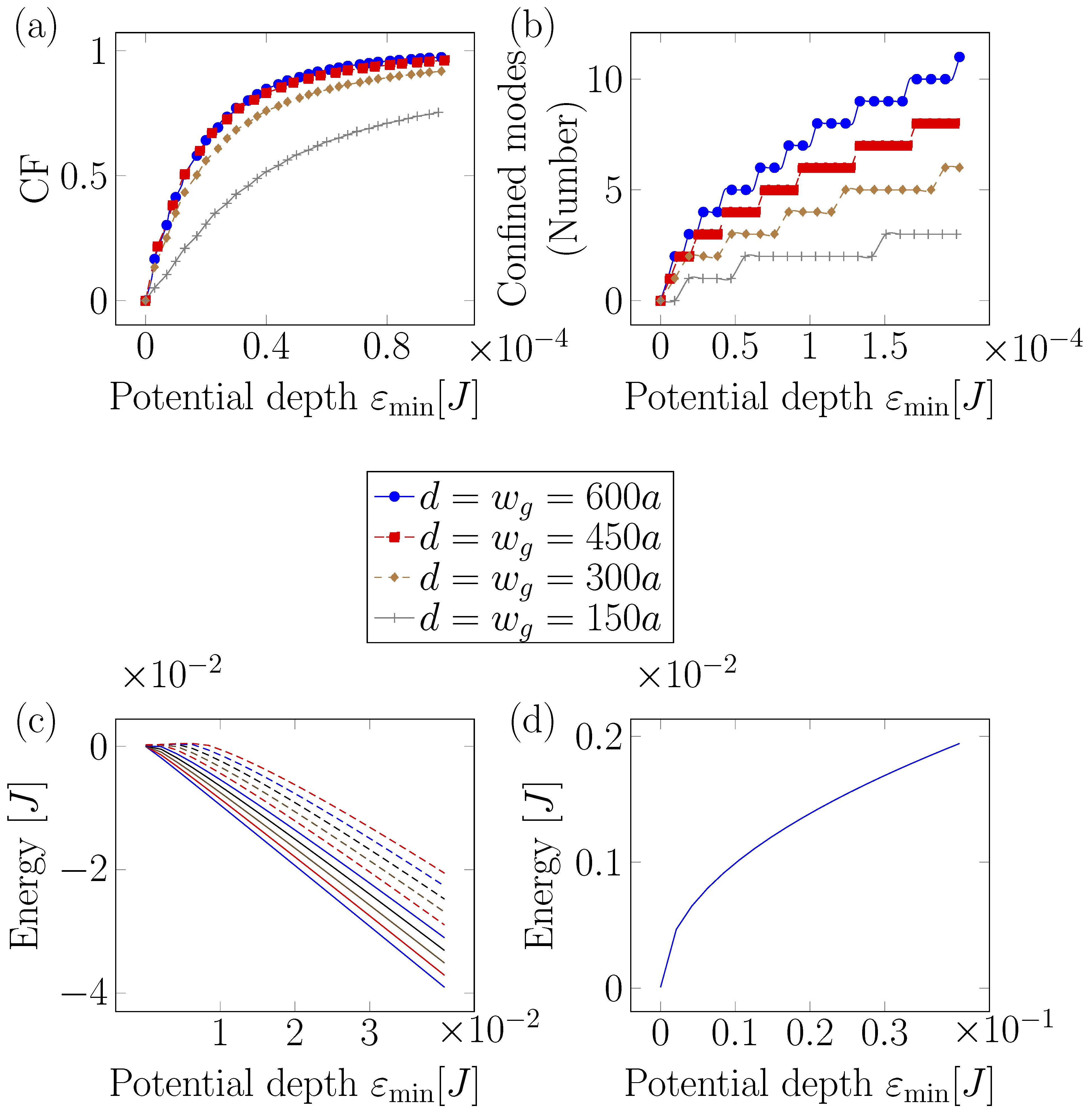} }
 \caption{(a) Confinement factor as a function of $\varepsilon_{\textrm{min}} [J]$ for several values of $d$. To achieve the same CF, a potential with smaller $d$ requires larger $\varepsilon_{\textrm{min}} [J]$. (b) Number of confined modes inside the potential.      
The criterion of confinement was chosen to be the value of CF. When CF of a particular mode reaches 0.9, we considered it a confined mode. A similar dependence on $d$ holds true here, where smaller $d$ needs a stronger potential to confine the same number of modes as a guide with larger $d$. (c) Ten lowest eigenvalues as a function of $\varepsilon_{\textrm{min}} [J], d= w_g= 150a$. A value below zero is a bound mode. As the potential depth increases, more and more modes become bound modes. (d) The difference between ground state and the first excited state. The difference also increases with the increasing depth of the potential.}
 \label{CF}
\end{figure*}

\Cref{shape_n_depth}(b) shows the z-component of the magnetic field inside the sheet due to the individual wires and their combination. The maximum depth occurs at the middle of the wires. \Cref{shape_n_depth}(c) shows the depth of the potential $(\varepsilon_{\textrm{min}})$ as a function of $w_g$ and $d$. The white line marks the maximum depth of the potential and hence defines an optimal choice of $w_g$ and $d$, which is  $w_g = d$, i.e. the distance of the wires from the spin sheet should be the half-width of the guide to achieve the maximum potential depth, $\varepsilon_{\textrm{min}}$. If the half-width, $w_g$, of the guide is significantly larger than its distance from the spin sheet, $d$, then the potential splits into two separate potentials (inset \cref{shape_n_depth}). The region of split potential is marked in black in \cref{shape_n_depth}(b). On the other hand if the spin-sheet to guide distance, $d$, is much larger then its half-width, $w_g$, this results in a shallow potential from which the magnon can escape more easily. Therefore, we restrict ourself to the choice of geometry where $w_g =  d$. Substituting this and  $x=0$ in \cref{f_form}, we get an expression for the potential depth $B_z^{\textrm{min}} = \mu I / 2\pi d$. The energy of a spin due to the $B_z^{\textrm{min}}$, which corresponds to the centre of the guide, see \cref{shape_n_depth} (a), is $\varepsilon_{\textrm{min}} = \gamma \hbar B_z^{\textrm{min}} = \gamma \hbar \mu I / (2 \pi d) = I/(\kappa d)$, where $\gamma$ is the gyromagnetic ratio, $\mu$ is the magnetic permeability of the material and $ \kappa = 2\pi/(\gamma\hbar\mu)$. This expression gives a clear connection between geometry, energy and the current in the wires.

For the rest of the paper we will use the potential depth $\varepsilon_{\textrm{min}}$ as the independent variable. For any given scale $d = w_g$, the potential depth is proportional to current $I$. The required $\varepsilon_{\textrm{min}}$ to achieve a certain confinement profile scales linearly with the strength of the $J$-coupling and therefore $\varepsilon_{\textrm{min}}$ is represented in units of $J$. With the knowledge of $J$ of a particular system and the geometry of the guide, one can calculate the current that is required in the wires to produce the required $\varepsilon_{\textrm{min}}$.      
 
Our spin guide model provides confinement analogous to that in optical wave guides and it can show similar functionalities, for example,  bending and coupling. Although the shape will have an effect on the properties like adiabatic bending and coupling, which we discuss in detail in the next section, the exact shape of the confining potential is not important for the guiding. As long as the potential is strong enough to create one or more confined modes, guiding can be achieved.

\section{Confinement factor and bend loss}
\label{sec-3}

Confinement factor is a commonly used term in semiconductor laser physics, which is defined as the ratio of the electric field in the active region to the total electric field \cite{Huang:1996fg}. We use the same analogy and define a magnonic version of this confinement factor based on the probability distribution of the magnon confined modes, which, in our case is the population inside the guide.

 \begin{figure*}[ht!]
   \centering
  \resizebox{1.8\columnwidth}{!}{
  \includegraphics{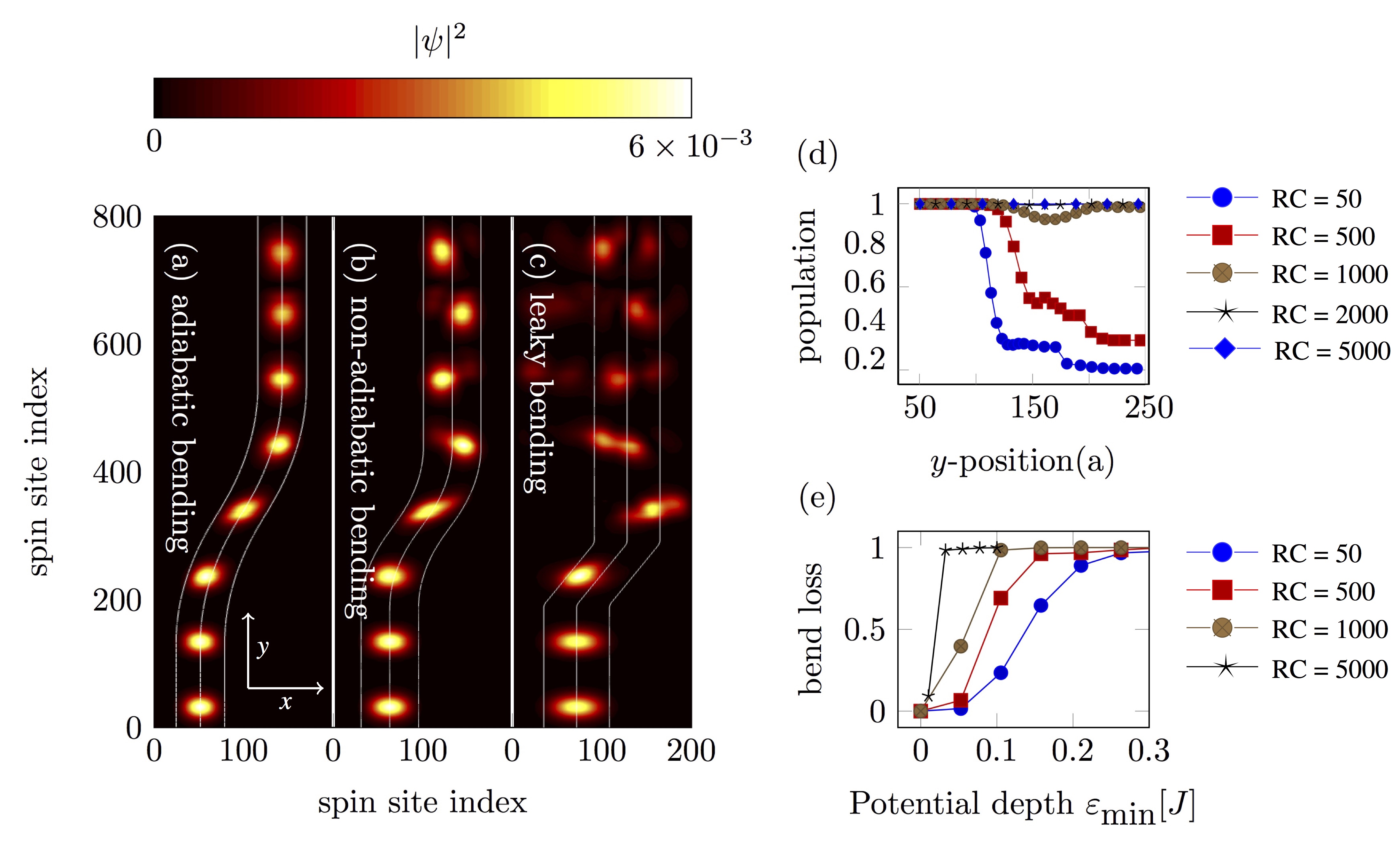}
};

      \caption{Effect of guide bending on the quality of transport. Subfigure (a)-(c) shows overlaid snapshots of equally spaced time-evolved states, where RC is the radius of curvature of the bend and the color axis is $|\psi|^2$. (a) Adiabatic bending, in which the magnon stays in the transversal ground state throughout the propagation. (b) As we make the bend tighter, the magnon starts to excite higher modes, although, it still stays confined in the guide. (c) At small bend radius the magnon leaks out of the guide. (d) Population inside waveguide as a function of $y$-position, for several radii of curvatures. (e) Bend loss as a function of potential depth, $\varepsilon_{\textrm{min}}$ for several radii of curvature. For a sharper bend, a deeper well is required to avoid bend loss.}
\label{bloss}
    \end{figure*}

   \begin{figure*}[ht!]
     \centering
   \resizebox{1.0\columnwidth}{!}{
\includegraphics{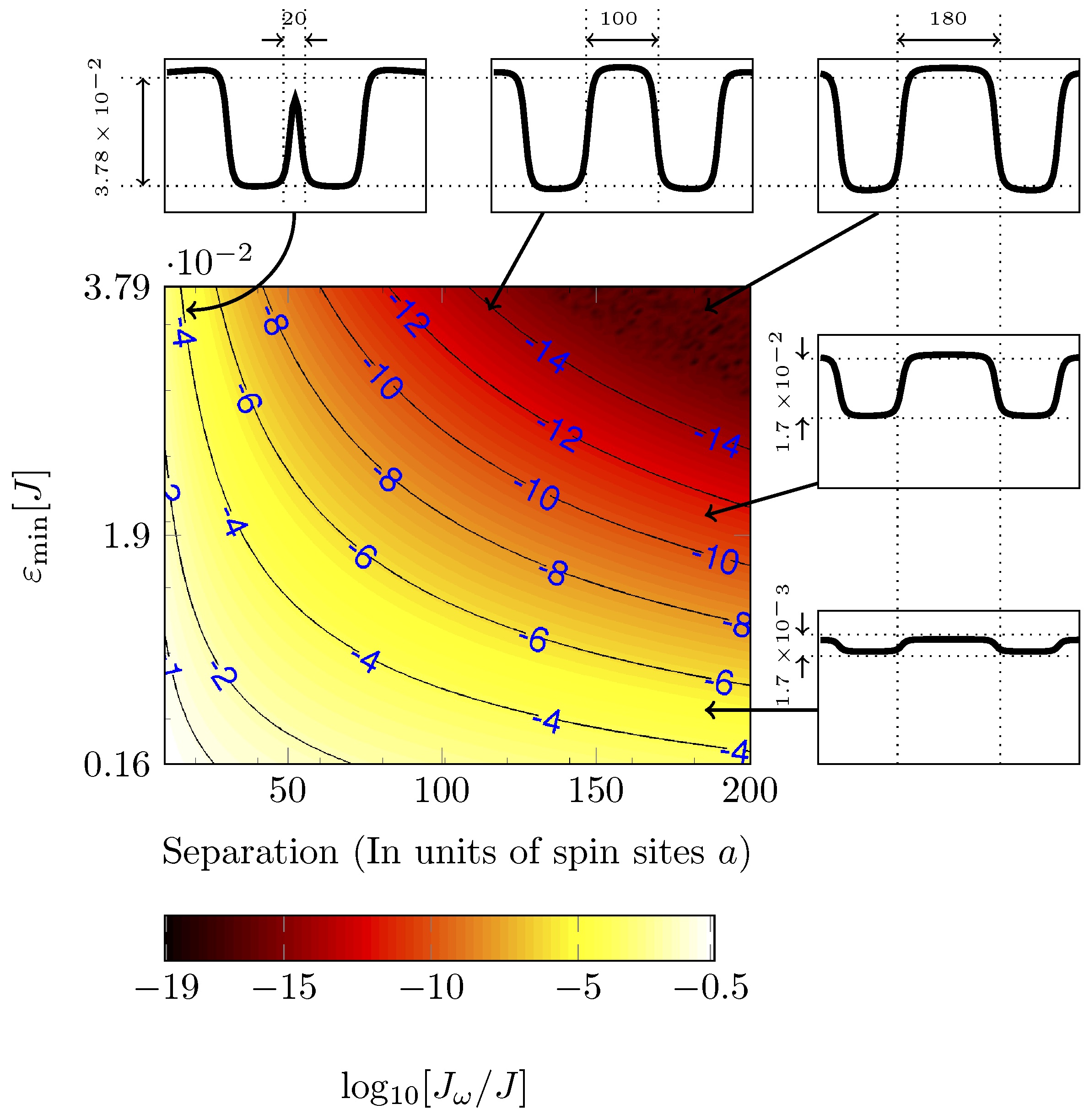}}
    \caption{Contours of constant coupling $J_\omega$ between two spin-guides acting as a directional coupler, as a function of the potential depth and the separation between the guides. The coupling strength decreases with increasing depth as well as with increasing separation between guides. At small currents, the modes become unbound. The thumbnails on the top and right sides show cross sections of the magnetic field perpendicular to the guiding direction, to illustrate the relative change in the shape of the potential along both axes.}
    \label{cont}
    \end{figure*}

\begin{align}
\centering
\textrm{CF} = \int^{w_g}_{-w_g}|\psi_{n}|^2 
%{1 - \int^{w_g}_{-w_g}|\psi_{n}|^2},
\end{align}
 % bend loss
\noindent  
where $\psi_n$ is the transversal wavefunction of the $n$-th eigen mode of magnon in the presence of the potential. CF varies between 0 an 1 and gives a measure of spacial confinement. 
\Cref{CF}(a) shows the confinement factor as a function of $\varepsilon_{\textrm{min}}$ for several values of $d$. A guide with a smaller width requires a stronger potential as compared to a guide with large width, to achieve the same value of CF.  \Cref{CF}(b) shows the number of confined modes for a potential as a function of $\varepsilon_{\textrm{min}} [J]$, for several widths. As with multi-mode optical fibers, the number of bound modes increases with both width and depth of the potential. Along with the spacial landscape, the energy landscape also changes with the changing potential. As the potential is applied, eigen modes lower their energies to become bound modes. \Cref{CF}(c) shows the first ten transversal eigen values as a function of $\varepsilon_{\textrm{min}}$. As expected, energies decrease with the increasing depth of the potential. \Cref{CF}(d) shows the energy difference between the ground state and the first excited state. The energy levels also grow apart as they transit from unbound to bound modes.

Similar to an optical waveguide, we can ``bend" these guides up to a certain bend radius without losing any confinement\cite{Clausen:2011fe,Vogt:2012ct}. We show some examples of bending in \cref{bloss}, where we identify three different bending regimes. When the radius of curvature of the bend is large, the magnon can follow the guide adiabatically, \cref{bloss}(a). As the radius of curvature reduces, the magnon will leak to higher confined modes, \cref{bloss}(b). In this kind of bending, the quantum phase is not preserved, so it is not a useful regime for quantum transport. Finally, when the bend becomes very sharp (small radius of curvature), the magnon cannot follow the guide anymore and it leaks out into unconfined modes,  \cref{bloss}(c). \Cref{bloss}(d) shows the population inside the guide, for several radii of curvature, as a function of distance along the propagation direction. The bending starts at the 100th site. Radii of curvature of 50$a$ and 500$a$ result in partial leakage of the population. The radius of curvature of 1000$a$ is a non-adiabatic bending in which population oscillates inside the guide but does not leak out. Finally the radii of curvatures of 2000$a$ and 5000$a$ show adiabatic transport. 

    \begin{figure}[t]
    \centering
    \resizebox{\columnwidth}{!}{
\includegraphics {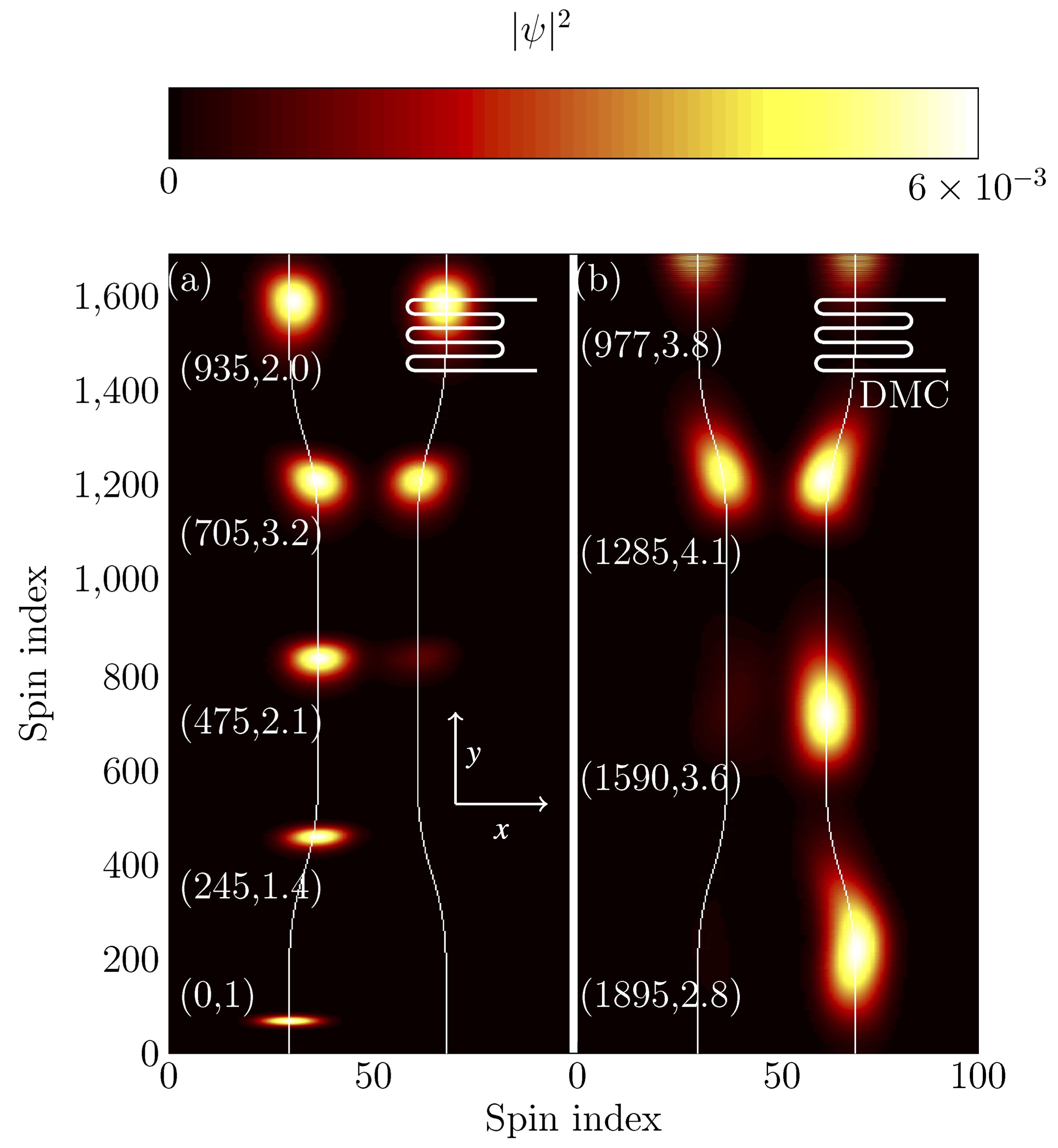}}
       \caption{Instances of full time evolution of a Michelson interferometer (for the device schematics see \cref{F1}(b) ). White lines represent the centre of each guide. Snap shots of the population are overlaid on top of each other, where the colour represents the $|\psi|^2$. Each snapshot is scaled such that its peak value appears as bright as the first instance. The right arm has a DMC at its end, which is used as a tuneable source of phase difference between both arms (see \cref{p_shift} for detail). Numbers placed next to each snap shot are the time stamps (in units of $1/J$) and scaling factor respectively. (a) The magnon was initialised in state $|L\rangle$ and it splits into  $(1/\sqrt{2})(|L\rangle + |R\rangle)$ upon passing through the splitter. In this particular case there is no relative phase shift upon reflection.  (b) After reflection and second pass through the splitter the magnon is transmitted to $|R\rangle$.}
    \label{mint_2}
    \end{figure}
\begin{figure}[h!]
\centering
\resizebox{\columnwidth}{!}{
\includegraphics[scale = 0.1]{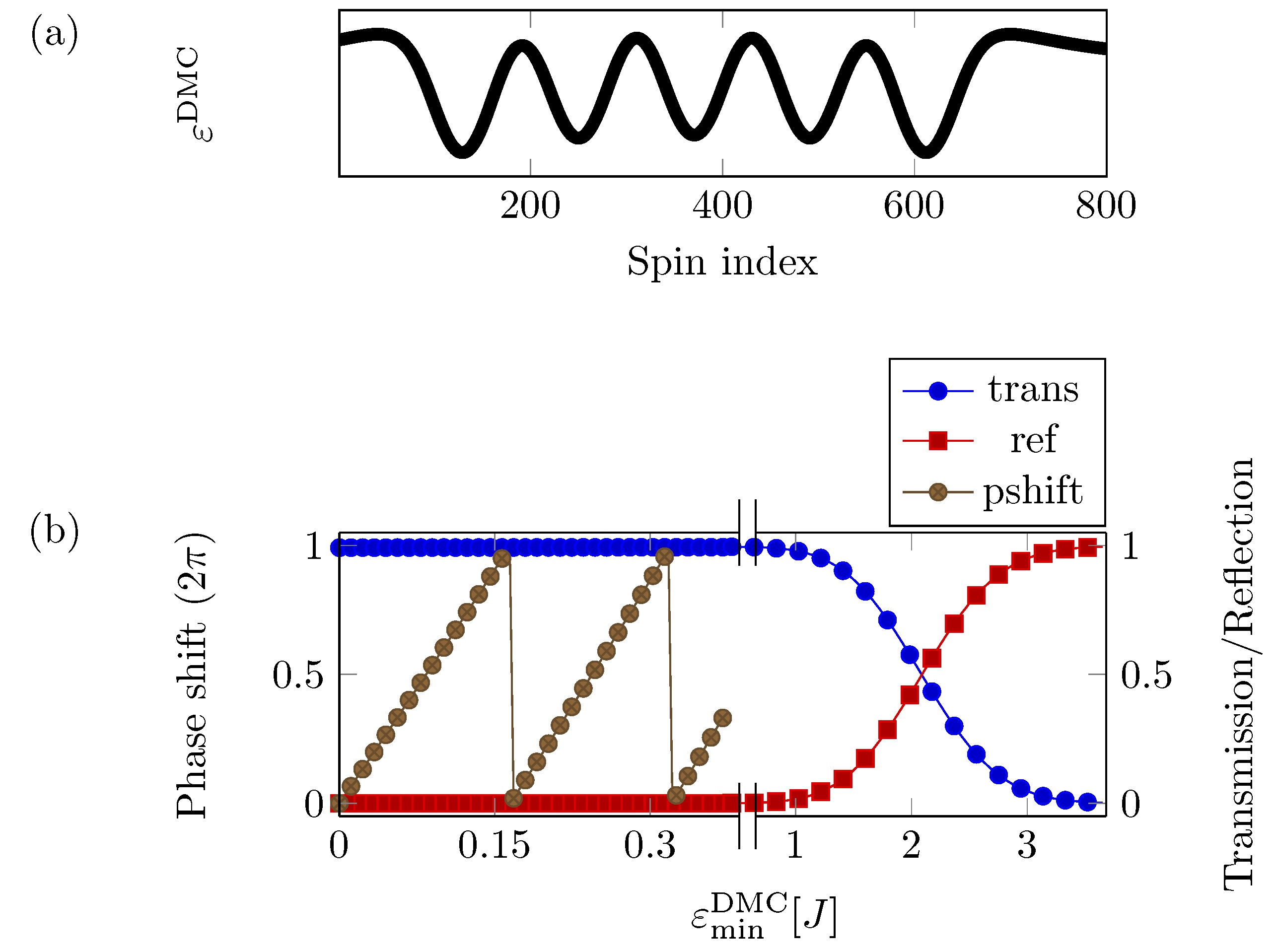}};

       \caption{(a) The $z$-component of the magnetic field from a DMC. DMC was modelled with 10 periodic repetitions with inter-wire separation of 20 spin sites. It takes two wires to produce one period of the magnetic field. (b) Phase shift as a function of current in DMC. When the $\varepsilon_{\min}^{\textrm{DMC}}$ field is greater then $J$, it acts as a reflecting potential and is no longer useful for phase control. Note the break in the horizontal scale.}
  \label{p_shift}
 \end{figure}

\section{Magnon splitter, and the Michelson interferometer}

    We now address the issue of the design of a magnonic Michelson interferometer.  The Michelson interferometer comprises an input channel, a splitter element (beamsplitter or directional coupler), a variable (ideally tuneable) phase element, and a reflective element that directs the signal back to the splitter.  The final output path of the magnon then becomes a sensitive function of the variable phase.  We discuss these elements in turn.
    
    The directional coupler is an important and commonly used device in optical fiber technology. Typically a directional coupler requires two proximal guides so that particles evanescently hop from one guide to the other.  Similarly, we can achieve the same functionality with our magnonic analogs.  When two guides are close to each other, the magnon can coherently tunnel between the guides and this tunnelling rate varies as a function of both the distance between the guides and the depth of the guides. If the guides are made lithographically, then the separation between the guides cannot be varied post-fabrication, however in our case the depth of the guides can be easily modified by varying the magnitude of the current in the surface wires. Hence we can realise reconfigurable directional couplers. 
     
  Figure~\ref{cont} shows  contours of constant coupling energy as a function of current and guide separation. The coupling is stronger when the guides are closer and shallower (lower current). To calculate the coupling energy, we performed a full Hamiltonian diagonalisation of a transversal cross section of the spin sheet to calculate the transversal ground state of each potential. Then the exchange energy was calculated as $\langle L |H| R \rangle$, where $ |L\rangle$ and $|R\rangle $ are the ground states of the left and the right guide and $H$ is the Hamiltonian with both guides in effect.

For a Michelson interferometer, we envisage adiabatically reducing the distance between the guides to effect the directional coupler, and then adiabatically increasing the distance after half of the population has been transferred, i.e. we perform the transformation $ |L\rangle \rightarrow (1/\sqrt{2})(|L\rangle + |R\rangle)$. The distance that is required to transfer half of the population is $l_{1/2} = v_g \pi a/2J_{\omega}$. 

In our model we started the guides at 35 sites apart and then adiabatically reduced the distance between them to 23 sites. The width of each guide was 20 sites, radius of the curvature of bend was 5000 sites and $\varepsilon_{\textrm{min}} $ was 1$J$. The coupling strength for this particular geometry was 0.0048$J$, which gives $l_{1/2} = 650$ sites, \cref{mint_2}(a).

 \begin{figure} [ht!]      
\centering
  \resizebox{\columnwidth}{!}{
 \includegraphics{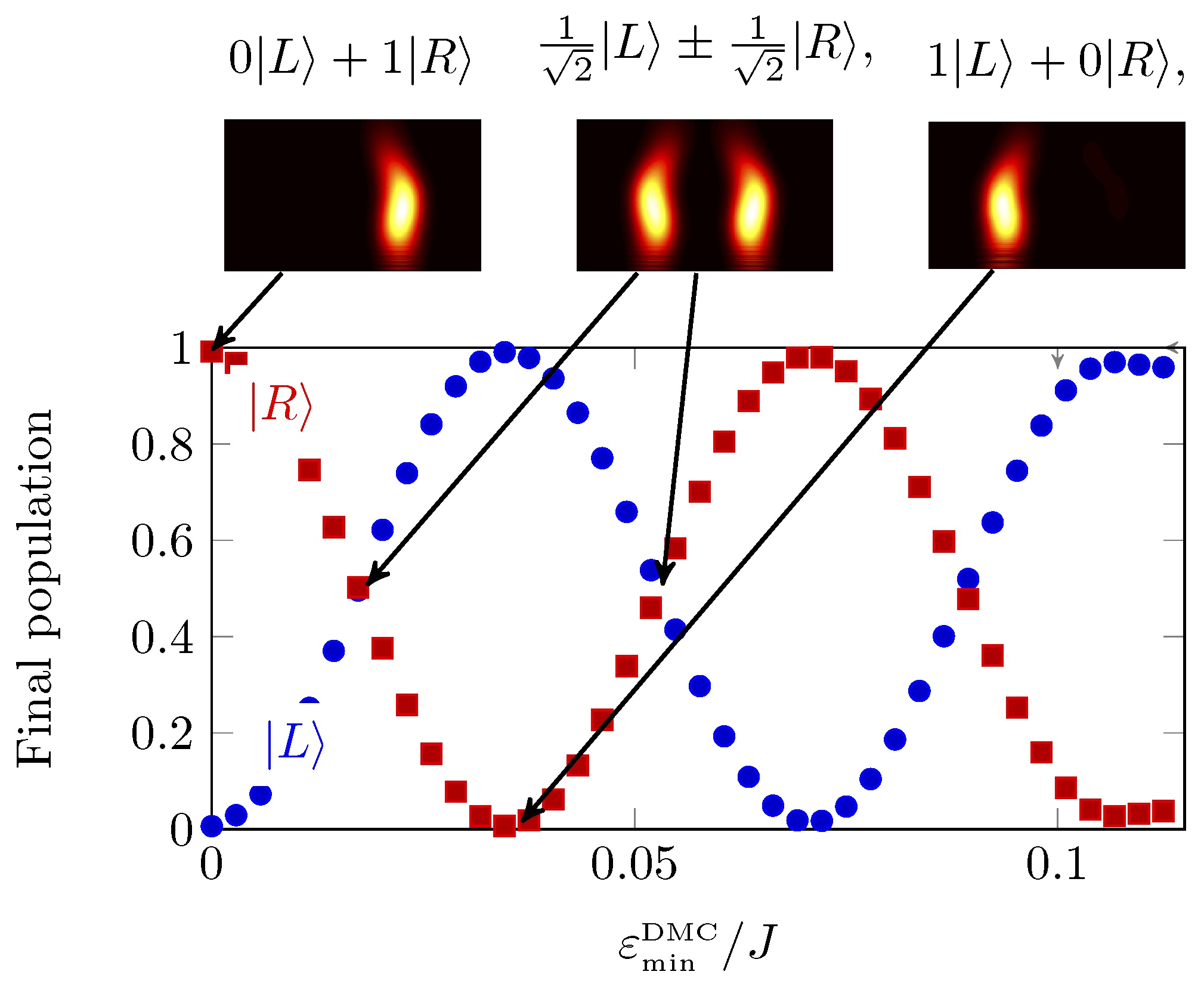}}
 \caption{Population in each arm at the end of the protocol, as a function of the depth of the potential due to the magnonic crystal, which in turn is a function of current in the crystal. The population in right and left arm is shown in red and blue. We initialise the magnon in state $|L\rangle$. With no phase shift the final state is $|R\rangle$. As the potential depth increases, the population varies sinusoidally between zero and one in both arms. The population goes through the whole state space and comes back to $|L\rangle$ at $\varepsilon^{\textrm{DMC}}_{\textrm{min}}/J \approx 0.07$.}
 \label{pop}
  \end{figure}
 
 \Cref{mint_2} shows the evolution where no extra phase ($\phi = 0$) is added to the magnon. \Cref{mint_2}(a) shows the propagation up to the reflection, whereas \cref{mint_2}(b) shows the propagation after the reflection.  The double passing of the beamsplitter has resulted in the magnon being shifted from $|L\rangle$ to $|R\rangle$. 
    
    There are many ways in which a controllable phase shift can be introduced.  Here we modelled a dynamic magnonic crystal-like structure \cite{Chumak:2009fz} at the end of one arm, as shown in \cref{mint_2}, which provides a tuneable phase shift in the right arm. A dynamic magnonic crystal is an array of equidistant conducting wires placed on the top of magnetic medium. The flow of current in these wires causes a periodic magnetic perturbation in the magnetic film as shown in \cref{p_shift} (a), which is analogous to a one dimensional magnetic crystal. The depth of the  perturbation is controlled by the current in the wires \cite{Krawczyk:2014ez}. In our case the periodicity of the magnonic crystal was chosen to be 20 spin sites with 10 periods. As we change the current in the crystal, we change the depth of the potential wells caused by the DMC. The phase change picked up by the magnon will be given by the integral over the entire DMC potential, which is linear in the minimum  of the potential, $\varepsilon_{\textrm{min}}^{\textrm{DMC}}$ [\cref{p_shift}(b)].

    \Cref{p_shift}(b)  also shows the reflection and transmission of the magnon from the DMC as a function of $\varepsilon_{\textrm{min}}^{\textrm{DMC}}$. The reflection from the DMC potential will be a function of magnon momentum. For a magnon moving with a speed $2aJ$, which is in the desired linear dispersion regime, we find for {$\varepsilon_{\textrm{min}}^{\textrm{DMC}}$}$<J$, there is no appreciable reflection of the magnon. As $\varepsilon^{\textrm{DMC}}_{\textrm{min}}$ becomes larger than $J$, it starts to reflect the magnon. Note that this reflection threshold depends on the speed of the magnon. Slower magnons will be reflected from weaker potentials. All of our calculations were done in the limit $\varepsilon_{\textrm{min}}^{\textrm{DMC}}\ll J$.

Reflection was achieved by the hard wall boundary condition at the edge of the spin sheet, which reverses the direction of the magnon, does not introduce any relative phase between the two arms of the interferometer.  After reflection, the population in the right arm passes the dynamic magnon crystal a second time, doubling the effect of its phase shift.     
For a given device, the final state of the interferometer is only a function of the relative phase between two arms, which in turn is a function of current in the crystal.
    
Considering the case where the magnon always initialised in $|L\rangle$, increasing the  current in the DMC leads to a relative phase shift in the right arm.  This leads to a sinusoidal variation in the final population in each arm as a function of the phase shift, as expected (\cref{pop}). When  $\varepsilon^{\textrm{DMC}}_{\textrm{min}}/J \approx 0.017$ the final population is equal in each arm ($1/\sqrt{2})(|L\rangle + |R\rangle$). As we keep increasing the current, the trend continues and the final state goes through $|L\rangle$, $(1/\sqrt{2})(|L\rangle - |R\rangle)$ and eventually coming back to $|R\rangle$ at $\varepsilon^{\textrm{DMC}}_{\textrm{min}}/J \approx 0.07$. This shows that the current through the dynamic magnonic crystal can be used as a tuneable element for magnon switching. 

\section{Realistic systems}
Up until now we have presented our results for a generic Hamiltonian. To translate our scheme to any particular implementation only requires knowledge of the coupling strength $J$, spin separation $a$ and the desired wire to spin-sheet separation $d$.
Here we consider a two-dimensional spin sheet of nano-patterned phosphorous in a silicon lattice \cite{Schofield:2003kr}. This system has been widely studied due to its relevance to phosphorous in silicon quantum computation \cite{Kane:1998wh,Kettle:2004hn,Koiller:2004gs,Koiller:2001gw,Wellard:2004fb,Wellard:2003jq} and quantum transport \cite{Smith:2015fo,Drumm:2013bs,Greentree:2004di}.  

We take the inter-donor spacing to be $a $=10nm, for which the J-coupling between donors is 40$\mu$eV\cite{Wang:2016ha}. If we chose the spin guide dimensions $w_g = d =100a = 1\mu$m, then according to \cref{cont} we require $\varepsilon_{\textrm{min}} = 1 \times 10^{-4} J$ to confine a single mode. The current required to produce that onsite energy is $I/d\kappa = \varepsilon_{\textrm{min}}$, where $\kappa = 2\pi/\gamma\hbar\mu$, which gives $I$ =  2.74 $\mu$A.

For the Michelson interferometer we need two further geometrical parameters; the half-coupling length -- the length that is required to transfer half of the magnon population to the other guide -- and the adiabatic radius of curvature, which is the radius at which magnon turns without leaking or coupling to the other modes. For a high contrast Michelson interferometer we chose $l_{1/2} = 500a$ which correspond to an inter-guide separation of $50a = 0.5 \mu m$ ($l_{1/2} = v_g \pi a/2J_{\omega}$). For a bending guide stronger potential is required to keep the magnon confined. If the chosen radius of curvature for interferometer is 5000a, then the required $\varepsilon_{\textrm{min}}$ will be $10^{-3}J$ \cref{bloss}(e) which corresponds to a current of 27.4$\mu$A. These parameters are well within the experimentally achievable limits and would lead to a total device length of $5\mu m$. There is considerable flexibility in the choice of parameters to build a working device.  

One important parameter to consider is the coherence time of the magnon. The magnon coherence time should be long enough to allow the magnon to complete the return trip through the interferometer  with sufficient fidelity to observe the interference effects. In the Heisenberg framework the maximum speed of the magnon is $2Ja/\hbar$,  which gives 607m/s for this particular system. This results in 16.7ns round trip time through a device of length 5$\mu$m. The most recently reported T-2 in P:Si is 2s at 5K temperature \cite{Tyryshkin:2012fi}, which is orders of magnitude longer than the magnon travel time.

    \section{Conclusion}
    We have proposed a scheme for guiding magnons in a two-dimensional ferromagnetic sheet using surface current carrying wires. Through numerical simulation based on the Heisenberg model, we demonstrated that transversal confinement can be achieved in this setup. We also presented a model of a magnonic Michelson interferometer. The magnon is split and recombined using a magnonic equivalent of a directional coupler. The extra phase was added onto one arm through a dynamical magnonic crystal. By changing the amount of current in the crystal one can obtain any desired combination of population in each arm. Equally, this device could be used as a sensor of any field capable of perturbing the acquired magnon phase.

    One possibility afforded by our scheme is the design of magnonic devices capable of performing non-determinstic linear magnonic quantum computation, by analogy with non-deterministic linear optical quantum computation \cite{Knill:2001is}.  Non-determinstic linear schemes utilise interferometric elements and the `hidden' nonlinearity introduced by measurement.  When magnons are distributed over many spins, as we have considered here, than they behave as simple type II bosons, and therefore show bunching in Hong-Ou-Mandel type configurations \cite{Hong:1987gm}.  Therefore  our results imply that full non-deterministic quantum gate operations can be simply ported to the magnonic case.  One further advantage of our reconfigurable scheme is that it is possible to dynamically switch guides on and off, and this may lead a natural realisation of schemes with `shortcuts' through high dimensional spaces \cite{Lanyon:2008gv}, and more generally to non-trivial consideration of optimisation of the Hilbert-space dimensionality of the resulting magnonic circuit for optimal computation \cite{Greentree:2004im}.
  \clearpage

 \section{appendices}
 \subsection{Effect of diagonal coupling}

 \begin{figure} [ht!]      
\centering
  \resizebox{\columnwidth}{!}{
 \includegraphics{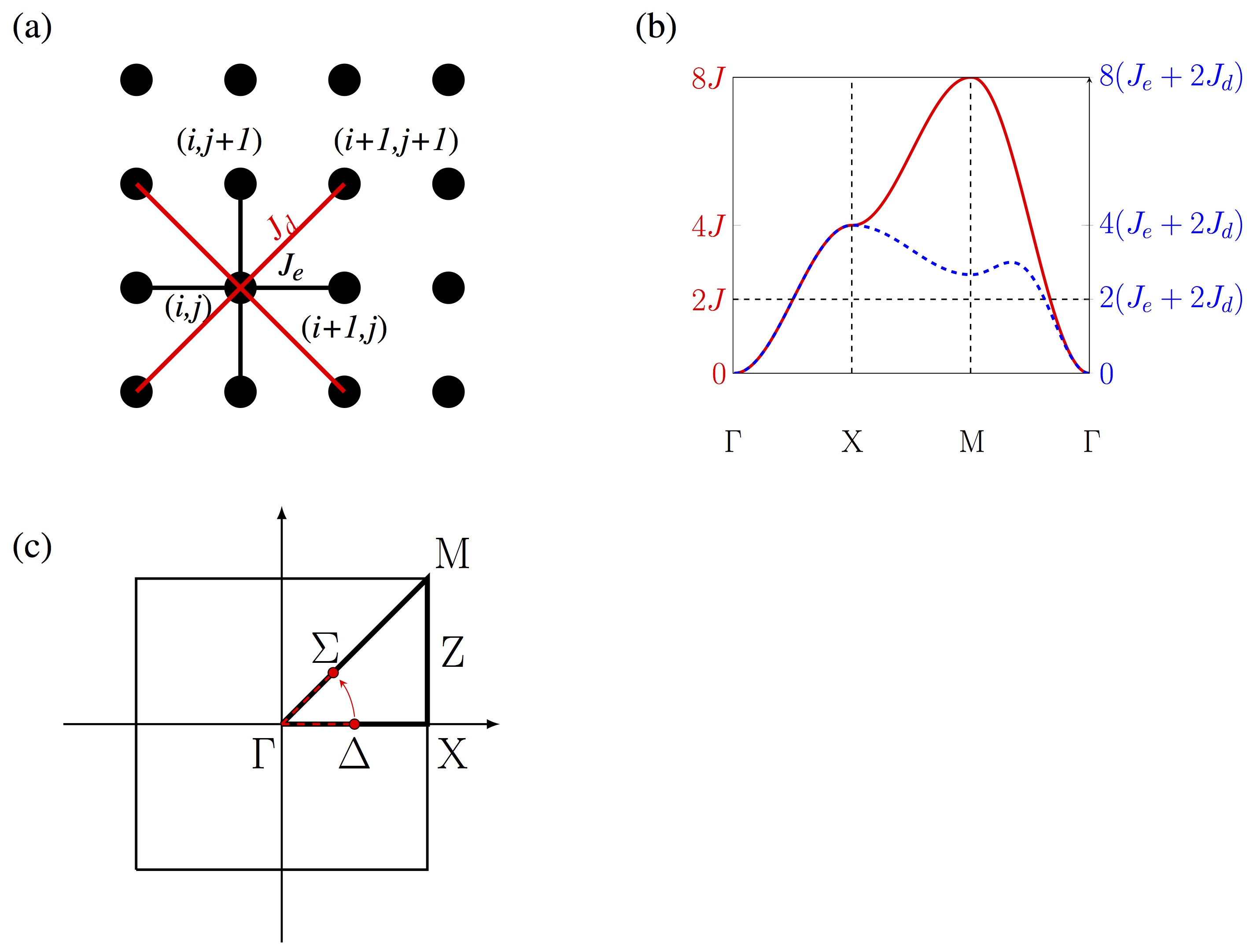}}
  \caption{(a) A Square lattice, where the black dots represent the lattice sites.  The black lines is the standard nearest-neighbour coupling $J_e$ i.e.\ the coupling between $(i,j)$ and $(i \pm1,j)$ , and between  $(i,j)$  and $(i,j \pm1)$. The red lines represent the next-nearest-neighbour or the diagonal couplings $J_d$, which is the coupling between $(i,j)$ , $(i\pm1,j\pm1)$. (b) Dispersion relation of a two-dimensional sheet along the high symmetry lines, with (blue-dotted) and without (red) the diagonal couplings ($J_d$). Along the $\Gamma$-X-direction, which corresponds to the principle axis in real space, both lines coincide. However, they differ completely along X-M arm and partially along the M-$\Gamma$ arm of the dispersion relation. (c) A two dimensional k-space of a spin-sheet. Two red dots and the arrow indicated the initial and final position of a magnon that was launched along the principle axis and then bent to 45 degrees.}
\end{figure}

Here we will compare the two Hamiltonians by calculating their dispersion relation through their action on an arbitrary wavefunction. A two dimensional Heisenberg Hamiltonian with the diagonal coupling is,

 \begin{align}
    H&=-J_e \sum_{i,j} \big[ {\mathbf{S}}_{i,j}.{\mathbf{S}}_{i+1,j} +{\mathbf{S}}_{i,j}.{\mathbf{S}}_{i,j+1} \big] \\  \label{diag_ham}  &-J_d \sum_{i,j} \big[{\mathbf{S}}_{i,j}.{\mathbf{S}}_{i+1,j+1}  + {\mathbf{S}}_{i,j}.{\mathbf{S}}_{i+1,j-1} \big] + \sum_{i,j} \varepsilon_{(i,j)},\nonumber \ \     
\end{align} 
where $J_e$ is the standard edge coupling and $J_d$ is the diagonal coupling. 

If $\psi$ is the eigen-state of the \cref{diag_ham} i.e. $H|\psi> = E |\psi> $, where E is the corresponding eigen-energy, then, expanding the left side of this equation into the action of the \cref{diag_ham} on the state $\psi$ ,

\begin {align}
& ( \psi_{i-1,j} + \psi_{i,j-1} + \psi_{i+1,j} + \psi_{i,j+1}- 4\psi_{i,j}) +  \nonumber \\
&  \frac{J_d}{J_e}( \psi_{i-1,j-1} + \psi_{i-1,j+1}  +\psi_{i+1,j-1} + \psi_{i+1,j+1} - 4\psi_{i,j})   \nonumber \\ 
& = -\frac{E}{J_e}\psi_{i,j} .
\end{align}
$\psi_{m,n}$ is the $(m,n)$th element of the eigne-vector, where $m$ and $n$ corresponds to the spin site indices in $x$ and $y$-direction.   

Using a trial wavefunction $\psi_{(m,n)} = e^{i(k_x+k_y)a}$ above equation becomes

\begin{align}
&(e^{i[k_x(i-a)+k_yj]} + e^{i[k_x(i+a)+k_yj]} +e^{i[k_xi+k_y(j-a)]}  \nonumber \\
&+ e^{i[k_xi+k_y(j +a)]}-4) +\frac{J_d}{J_e} (e^{i[k_x(i-a)+k_y(j-a)]} \nonumber \\
&+ e^{i[k_x(i-a)+k_y(j +a)]}+e^{i[k_x(i+a)+k_y(j-a)]}  \nonumber \\
& + e^{i[k_x(i+a)+k_y(j +a)]} -4) = \frac{-E}{J_e}e^{i(k_xi+k_yj)} 
\end{align}

\begin{align}
   &(e^{-ik_xa} + e^{ik_xa} + e^{-ik_ya} + e^{ik_ya} - 4) \nonumber \\ 
   &+  \frac{J_d}{J_e}(e^{-i(k_xa+k_ya)} + e^{-i(k_xa-k_ya)} \nonumber \\
   &+ e^{i(k_xa-k_ya)}   + e^{i(k_xa+k_ya)} - 4)  = \frac{-E}{J_e} 
\end{align}

\begin{align}
\frac{-E}{J_e}&  =   2[\textrm{cos}(k_xa) + \textrm{cos}(k_ya)  + 2] \nonumber \\
 &+ 2\frac{J_d}{J_e}[\textrm{cos}(k_xa+k_ya) + \textrm{cos}(k_xa-k_ya) + 2] 
\end{align}

\begin{align}
E =\ &4(J_e+J_d) - 2J_e[\textrm{cos}(k_xa) + \textrm{cos}(k_ya)] \nonumber \\
&- 4J_d[\textrm{cos}(k_xa) \textrm{cos}(k_ya)]
\end{align}

\noindent If the magnon is only traveling along one direction, say in x-direction and $k_y = 0$, then the above equation becomes

\begin{align}
E = &  2(J_e+2J_d)[1-\textrm{cos}(k_xa)] 
\end{align}

This is exactly a dispersion relation of a spin-sheet except $J_e$ is now $J_e+2J_d$. 
\\

Figure 10 (a) shows a square lattice with lines representing the couplings. Black lines represent the edge coupling and red lines represent the diagonal couplings. Figure 10 (b) shows the dispersion relation of the spin sheet, with (blue-dotted) and without (red) the diagonal coupling. Both lines coincide for the $\Gamma$-X part of the curve, however the y-axis scales are different for both curves. This shows that the addition of $J_d$ acts as a scaling factor in the dynamics of a magnon that is traveling along the principle direction in a sheet with nearest-neighbour coupling. \\

The region of interest in the band structure is the region that is around the $2J$ line. This is due to the fact that we always launch the magnon with the energy of $2J$, which corresponds to the maximum gradient and hence maximum group velocity, which the magnons travel at. Once launched along the principle axis with the wave vector of $\pi/2$, which corresponds to the energy of 2$J$, any bending starts to transfer the population along the $\textrm{M}-\Gamma$ arm of the dispersion curve at $k = \pi/3$, while keeping the energy constant  [Fig 10(b,c)]. The $\textrm{M} - \Gamma$ arm of the dispersion relation, with and without the diagonal couplings, only differs by approximately $ 3\%$ at the energy of 2$J$.  Based on this we can approximate the dynamics of the magnon in the spin-sheet with a diagonal coupling by simply simulating a spin-sheet with a rescaled edge coupling.

 \acknowledgements
    The authors would like to thank Jared Cole and Ivan Maksymov, for useful discussions. This work was supported by the Australian Research Council (Grant No. DP130104381).

\bibliography{myref}

\begin{thebibliography}{10}
\expandafter\ifx\csname url\endcsname\relax
  \def\url#1{\texttt{#1}}\fi
\expandafter\ifx\csname urlprefix\endcsname\relax\def\urlprefix{URL }\fi
\providecommand{\bibinfo}[2]{#2}
\providecommand{\eprint}[2][]{\url{#2}}

\bibitem{Kruglyak:2010cy}
\bibinfo{author}{Kruglyak, V.~V.}, \bibinfo{author}{Demokritov, S.~O.} \&
  \bibinfo{author}{Grundler, D.}
\newblock \bibinfo{title}{{Magnonics}}.
\newblock \emph{\bibinfo{journal}{Journal of Physics D: Applied Physics}}
  \textbf{\bibinfo{volume}{43}}, \bibinfo{pages}{264001}
  (\bibinfo{year}{2010}).

\bibitem{Lenk:2011el}
\bibinfo{author}{Lenk, B.}, \bibinfo{author}{Ulrichs, H.},
  \bibinfo{author}{Garbs, F.} \& \bibinfo{author}{M{\"u}nzenberg, M.}
\newblock \bibinfo{title}{{The building blocks of magnonics}}.
\newblock \emph{\bibinfo{journal}{Physics Reports}}
  \textbf{\bibinfo{volume}{507}}, \bibinfo{pages}{107--136}
  (\bibinfo{year}{2011}).

\bibitem{Chumak:2015fa}
\bibinfo{author}{Chumak, A.~V.}, \bibinfo{author}{Vasyuchka, V.~I.},
  \bibinfo{author}{Serga, A.~A.} \& \bibinfo{author}{Hillebrands, B.}
\newblock \bibinfo{title}{{Magnon spintronics}}.
\newblock \emph{\bibinfo{journal}{Nature Physics}}
  \textbf{\bibinfo{volume}{11}}, \bibinfo{pages}{453--461}
  (\bibinfo{year}{2015}).

\bibitem{Khitun:2001bva}
\bibinfo{author}{Khitun, A.}, \bibinfo{author}{Ostroumov, R.} \&
  \bibinfo{author}{Wang, K.~L.}
\newblock \bibinfo{title}{{Spin-wave utilization in a quantum computer}}.
\newblock \emph{\bibinfo{journal}{Physical Review A}}
  \textbf{\bibinfo{volume}{64}}, \bibinfo{pages}{062304}
  (\bibinfo{year}{2001}).

\bibitem{Chumak:2014iw}
\bibinfo{author}{Chumak, A.~V.}, \bibinfo{author}{Serga, A.~A.} \&
  \bibinfo{author}{Hillebrands, B.}
\newblock \bibinfo{title}{{Magnon transistor for all-magnon data processing}}.
\newblock \emph{\bibinfo{journal}{Nature Communications}}
  \textbf{\bibinfo{volume}{5}}, \bibinfo{pages}{4700} (\bibinfo{year}{2014}).

\bibitem{Khitun:cb}
\bibinfo{author}{Khitun, A.}, \bibinfo{author}{Bao, M.} \&
  \bibinfo{author}{Wang, K.~L.}
\newblock \bibinfo{title}{{Spin Wave Magnetic NanoFabric: A New Approach to
  Spin-Based Logic Circuitry}}.
\newblock \emph{\bibinfo{journal}{IEEE Transactions on Magnetics}}
  \textbf{\bibinfo{volume}{44}}, \bibinfo{pages}{2141--2152}.

\bibitem{Khitun:2010dx}
\bibinfo{author}{Khitun, A.}, \bibinfo{author}{Bao, M.} \&
  \bibinfo{author}{Wang, K.~L.}
\newblock \bibinfo{title}{{Magnonic logic circuits}}.
\newblock \emph{\bibinfo{journal}{Journal of Physics D: Applied Physics}}
  \textbf{\bibinfo{volume}{43}}, \bibinfo{pages}{264005}
  (\bibinfo{year}{2010}).

\bibitem{Vogt:2014he}
\bibinfo{author}{Vogt, K.} \emph{et~al.}
\newblock \bibinfo{title}{{Realization of a spin-wave multiplexer}}.
\newblock \emph{\bibinfo{journal}{Nature Communications}}
  \textbf{\bibinfo{volume}{5}} (\bibinfo{year}{2014}).

\bibitem{Bauer:2012fq}
\bibinfo{author}{Bauer, G. E.~W.}, \bibinfo{author}{Saitoh, E.} \&
  \bibinfo{author}{van Wees, B.~J.}
\newblock \bibinfo{title}{{Spin caloritronics}}.
\newblock \emph{\bibinfo{journal}{Nature Materials}}
  \textbf{\bibinfo{volume}{11}}, \bibinfo{pages}{391--399}
  (\bibinfo{year}{2012}).

\bibitem{Bose:2007ji}
\bibinfo{author}{Bose, S.}
\newblock \bibinfo{title}{{Quantum communication through spin chain dynamics:
  an introductory overview}}.
\newblock \emph{\bibinfo{journal}{Contemporary Physics}}
  \textbf{\bibinfo{volume}{48}}, \bibinfo{pages}{13--30}
  (\bibinfo{year}{2007}).

\bibitem{Longo:2013kk}
\bibinfo{author}{Longo, P.}, \bibinfo{author}{Greentree, A.~D.},
  \bibinfo{author}{Busch, K.} \& \bibinfo{author}{Cole, J.~H.}
\newblock \bibinfo{title}{{Quantum Bocce: Magnon{\textendash}magnon collisions
  between propagating and bound states in 1D spin chains}}.
\newblock \emph{\bibinfo{journal}{Physics Letters A}}
  \textbf{\bibinfo{volume}{377}}, \bibinfo{pages}{1242--1249}
  (\bibinfo{year}{2013}).

\bibitem{Schneider:2008fu}
\bibinfo{author}{Schneider, T.} \emph{et~al.}
\newblock \bibinfo{title}{{Realization of spin-wave logic gates}}.
\newblock \emph{\bibinfo{journal}{Applied Physics Letters}}
  \textbf{\bibinfo{volume}{92}}, \bibinfo{pages}{022505}
  (\bibinfo{year}{2008}).

\bibitem{Serga:2009dg}
\bibinfo{author}{Serga, A.~A.}, \bibinfo{author}{Neumann, T.},
  \bibinfo{author}{Chumak, A.~V.} \& \bibinfo{author}{Hillebrands, B.}
\newblock \bibinfo{title}{{Generation of spin-wave pulse trains by
  current-controlled magnetic mirrors}}.
\newblock \emph{\bibinfo{journal}{Applied Physics Letters}}
  \textbf{\bibinfo{volume}{94}}, \bibinfo{pages}{112501}
  (\bibinfo{year}{2009}).

\bibitem{Chumak:2009fz}
\bibinfo{author}{Chumak, A.~V.}, \bibinfo{author}{Neumann, T.},
  \bibinfo{author}{Serga, A.~A.}, \bibinfo{author}{Hillebrands, B.} \&
  \bibinfo{author}{Kostylev, M.~P.}
\newblock \bibinfo{title}{{A current-controlled, dynamic magnonic crystal}}.
\newblock \emph{\bibinfo{journal}{Journal of Physics D: Applied Physics}}
  \textbf{\bibinfo{volume}{42}}, \bibinfo{pages}{205005}
  (\bibinfo{year}{2009}).

\bibitem{Neumann:2009hz}
\bibinfo{author}{Neumann, T.}, \bibinfo{author}{Serga, A.~A.} \&
  \bibinfo{author}{Hillebrands, B.}
\newblock \bibinfo{title}{{Frequency-dependent reflection of spin waves from a
  magnetic inhomogeneity induced by a surface direct current}}.
\newblock \emph{\bibinfo{journal}{Applied Physics Letters}}
  \textbf{\bibinfo{volume}{94}}, \bibinfo{pages}{042503}
  (\bibinfo{year}{2009}).

\bibitem{Chumak:2009eg}
\bibinfo{author}{Chumak, A.~V.}, \bibinfo{author}{Serga, A.~A.},
  \bibinfo{author}{Wolff, S.}, \bibinfo{author}{Hillebrands, B.} \&
  \bibinfo{author}{Kostylev, M.~P.}
\newblock \bibinfo{title}{{Scattering of surface and volume spin waves in a
  magnonic crystal}}.
\newblock \emph{\bibinfo{journal}{Applied Physics Letters}}
  \textbf{\bibinfo{volume}{94}}, \bibinfo{pages}{172511}
  (\bibinfo{year}{2009}).

\bibitem{Anonymous:f5yEVrch}
\bibinfo{author}{Gulyaev, Y.~V.} \emph{et~al.}
\newblock \bibinfo{title}{{Ferromagnetic films with magnon bandgap periodic
  structures: Magnon crystals}}.
\newblock \emph{\bibinfo{journal}{Jetp Letters}} \textbf{\bibinfo{volume}{77}},
  \bibinfo{pages}{567--570} (\bibinfo{year}{2003}).

\bibitem{Chumak:2009koa}
\bibinfo{author}{Chumak, A.~V.} \emph{et~al.}
\newblock \bibinfo{title}{{Spin-wave propagation in a microstructured magnonic
  crystal}}.
\newblock \emph{\bibinfo{journal}{Applied Physics Letters}}
  \textbf{\bibinfo{volume}{95}}, \bibinfo{pages}{262508}
  (\bibinfo{year}{2009}).

\bibitem{Vasiliev:2007br}
\bibinfo{author}{Vasiliev, S.~V.}, \bibinfo{author}{Kruglyak, V.~V.},
  \bibinfo{author}{Sokolovskii, M.~L.} \& \bibinfo{author}{Kuchko, A.~N.}
\newblock \bibinfo{title}{{Spin wave interferometer employing a local
  nonuniformity of the effective magnetic field}}.
\newblock \emph{\bibinfo{journal}{Journal of Applied Physics}}
  \textbf{\bibinfo{volume}{101}}, \bibinfo{pages}{113919}
  (\bibinfo{year}{2007}).

\bibitem{Kostylev:2005fua}
\bibinfo{author}{Kostylev, M.~P.}, \bibinfo{author}{Serga, A.~A.},
  \bibinfo{author}{Schneider, T.}, \bibinfo{author}{Leven, B.} \&
  \bibinfo{author}{Hillebrands, B.}
\newblock \bibinfo{title}{{Spin-wave logical gates}}.
\newblock \emph{\bibinfo{journal}{Applied Physics Letters}}
  \textbf{\bibinfo{volume}{87}}, \bibinfo{pages}{153501}
  (\bibinfo{year}{2005}).

\bibitem{Dyson:1956kb}
\bibinfo{author}{Dyson, F.~J.}
\newblock \bibinfo{title}{{General Theory of Spin-Wave Interactions}}.
\newblock \emph{\bibinfo{journal}{Phys.Rev.}} \textbf{\bibinfo{volume}{102}},
  \bibinfo{pages}{1217--1230} (\bibinfo{year}{1956}).

\bibitem{Heisenberg:1928jw}
\bibinfo{author}{Heisenberg, W.}
\newblock \bibinfo{title}{{Zur Theorie des Ferromagnetismus}}.
\newblock \emph{\bibinfo{journal}{Zeitschrift f{\"u}r Physik}}
  \textbf{\bibinfo{volume}{49}}, \bibinfo{pages}{619--636}
  (\bibinfo{year}{1928}).

\bibitem{Serga:2010cw}
\bibinfo{author}{Serga, A.~A.}, \bibinfo{author}{Chumak, A.~V.} \&
  \bibinfo{author}{Hillebrands, B.}
\newblock \bibinfo{title}{{YIG magnonics}}.
\newblock \emph{\bibinfo{journal}{Journal of Physics D: Applied Physics}}
  \textbf{\bibinfo{volume}{43}}, \bibinfo{pages}{264002}
  (\bibinfo{year}{2010}).

\bibitem{vanHoogdalem:2011bo}
\bibinfo{author}{van Hoogdalem, K.~A.} \& \bibinfo{author}{Loss, D.}
\newblock \bibinfo{title}{{Rectification of spin currents in spin chains}}.
\newblock \emph{\bibinfo{journal}{Physical Review B}}
  \textbf{\bibinfo{volume}{84}}, \bibinfo{pages}{024402}
  (\bibinfo{year}{2011}).

\bibitem{Jeske:2013cs}
\bibinfo{author}{Jeske, J.}, \bibinfo{author}{Vogt, N.} \&
  \bibinfo{author}{Cole, J.~H.}
\newblock \bibinfo{title}{{Excitation and state transfer through spin chains in
  the presence of spatially correlated noise}}.
\newblock \emph{\bibinfo{journal}{Physical Review A}}
  \textbf{\bibinfo{volume}{88}}, \bibinfo{pages}{062333}
  (\bibinfo{year}{2013}).

\bibitem{Makin:2012vi}
\bibinfo{author}{Makin, M.~I.}, \bibinfo{author}{Cole, J.~H.},
  \bibinfo{author}{Hill, C.~D.} \& \bibinfo{author}{Greentree, A.~D.}
\newblock \bibinfo{title}{{Spin Guides and Spin Splitters: Waveguide Analogies
  in One-Dimensional Spin Chains}}.
\newblock \emph{\bibinfo{journal}{Physical Review Letters}}
  \textbf{\bibinfo{volume}{108}}, \bibinfo{pages}{017207}
  (\bibinfo{year}{2012}).

\bibitem{Ahmed:2015eo}
\bibinfo{author}{Ahmed, M.~H.} \& \bibinfo{author}{Greentree, A.~D.}
\newblock \bibinfo{title}{{Guided magnon transport in spin chains: Transport
  speed and correcting for disorder}}.
\newblock \emph{\bibinfo{journal}{Physical Review A}}
  \textbf{\bibinfo{volume}{91}}, \bibinfo{pages}{022306}
  (\bibinfo{year}{2015}).

\bibitem{Nikolopoulos:2008ka}
\bibinfo{author}{Nikolopoulos, G.~M.}
\newblock \bibinfo{title}{{Directional Coupling for Quantum Computing and
  Communication}}.
\newblock \emph{\bibinfo{journal}{Physical Review Letters}}
  \textbf{\bibinfo{volume}{101}}, \bibinfo{pages}{200502}
  (\bibinfo{year}{2008}).

\bibitem{Yung:2005gm}
\bibinfo{author}{Yung, M.-H.} \& \bibinfo{author}{Bose, S.}
\newblock \bibinfo{title}{{Perfect state transfer, effective gates, and
  entanglement generation in engineered bosonic and fermionic networks}}.
\newblock \emph{\bibinfo{journal}{Physical Review A}}
  \textbf{\bibinfo{volume}{71}}, \bibinfo{pages}{032310}
  (\bibinfo{year}{2005}).

\bibitem{Makin:2009ja}
\bibinfo{author}{Makin, M.~I.}, \bibinfo{author}{Cole, J.~H.},
  \bibinfo{author}{Hill, C.~D.}, \bibinfo{author}{Greentree, A.~D.} \&
  \bibinfo{author}{Hollenberg, L. C.~L.}
\newblock \bibinfo{title}{{Time evolution of the one-dimensional
  Jaynes-Cummings-Hubbard Hamiltonian}}.
\newblock \emph{\bibinfo{journal}{Physical Review A}}
  \textbf{\bibinfo{volume}{80}}, \bibinfo{pages}{043842}
  (\bibinfo{year}{2009}).

\bibitem{Klingler:2014hk}
\bibinfo{author}{Klingler, S.} \emph{et~al.}
\newblock \bibinfo{title}{{Design of a spin-wave majority gate employing mode
  selection}}.
\newblock \emph{\bibinfo{journal}{Applied Physics Letters}}
  \textbf{\bibinfo{volume}{105}}, \bibinfo{pages}{152410}
  (\bibinfo{year}{2014}).

\bibitem{deSousa:2004dua}
\bibinfo{author}{de~Sousa, R.}, \bibinfo{author}{Delgado, J.~D.} \&
  \bibinfo{author}{Das~Sarma, S.}
\newblock \bibinfo{title}{{Silicon quantum computation based on magnetic
  dipolar coupling}}.
\newblock \emph{\bibinfo{journal}{Physical Review A}}
  \textbf{\bibinfo{volume}{70}}, \bibinfo{pages}{052304}
  (\bibinfo{year}{2004}).

\bibitem{Chumak:2015fab}
\bibinfo{author}{Chumak, A.~V.}, \bibinfo{author}{Vasyuchka, V.~I.},
  \bibinfo{author}{Serga, A.~A.} \& \bibinfo{author}{Hillebrands, B.}
\newblock \bibinfo{title}{{Magnon spintronics}}.
\newblock \emph{\bibinfo{journal}{Nature Physics}}
  \textbf{\bibinfo{volume}{11}}, \bibinfo{pages}{453--461}
  (\bibinfo{year}{2015}).

\bibitem{Demidov:2011jo}
\bibinfo{author}{Demidov, V.~E.} \emph{et~al.}
\newblock \bibinfo{title}{{Excitation of short-wavelength spin waves in
  magnonic waveguides}}.
\newblock \emph{\bibinfo{journal}{Applied Physics Letters}}
  \textbf{\bibinfo{volume}{99}}, \bibinfo{pages}{082507}
  (\bibinfo{year}{2011}).

\bibitem{Gieniusz:2013eb}
\bibinfo{author}{Gieniusz, R.} \emph{et~al.}
\newblock \bibinfo{title}{{Single antidot as a passive way to create caustic
  spin-wave beams in yttrium iron garnet films}}.
\newblock \emph{\bibinfo{journal}{Applied Physics Letters}}
  \textbf{\bibinfo{volume}{102}}, \bibinfo{pages}{102409}
  (\bibinfo{year}{2013}).

\bibitem{Huang:1996fg}
\bibinfo{author}{Huang, Y.-Z.}, \bibinfo{author}{Pan, Z.} \&
  \bibinfo{author}{Wu, R.-H.}
\newblock \bibinfo{title}{{Analysis of the optical confinement factor in
  semiconductor lasers}}.
\newblock \emph{\bibinfo{journal}{Journal of Applied Physics}}
  \textbf{\bibinfo{volume}{79}}, \bibinfo{pages}{3827} (\bibinfo{year}{1996}).

\bibitem{Clausen:2011fe}
\bibinfo{author}{Clausen, P.} \emph{et~al.}
\newblock \bibinfo{title}{{Mode conversion by symmetry breaking of propagating
  spin waves}}.
\newblock \emph{\bibinfo{journal}{Applied Physics Letters}}
  \textbf{\bibinfo{volume}{99}}, \bibinfo{pages}{162505}
  (\bibinfo{year}{2011}).

\bibitem{Vogt:2012ct}
\bibinfo{author}{Vogt, K.} \emph{et~al.}
\newblock \bibinfo{title}{{Spin waves turning a corner}}.
\newblock \emph{\bibinfo{journal}{Applied Physics Letters}}
  \textbf{\bibinfo{volume}{101}}, \bibinfo{pages}{042410--4}
  (\bibinfo{year}{2012}).

\bibitem{Krawczyk:2014ez}
\bibinfo{author}{Krawczyk, M.} \& \bibinfo{author}{Grundler, D.}
\newblock \bibinfo{title}{{Review and prospects of magnonic crystals~and
  devices with reprogrammable band structure}}.
\newblock \emph{\bibinfo{journal}{J. Phys.: Condens. Matter}}
  \textbf{\bibinfo{volume}{26}}, \bibinfo{pages}{123202}
  (\bibinfo{year}{2014}).

\bibitem{Schofield:2003kr}
\bibinfo{author}{Schofield, S.} \emph{et~al.}
\newblock \bibinfo{title}{{Atomically Precise Placement of Single Dopants in
  Si}}.
\newblock \emph{\bibinfo{journal}{Physical Review Letters}}
  \textbf{\bibinfo{volume}{91}}, \bibinfo{pages}{136104}
  (\bibinfo{year}{2003}).

\bibitem{Kane:1998wh}
\bibinfo{author}{Kane, B.~E.}
\newblock \bibinfo{title}{{A silicon-based nuclear spin quantum computer}}.
\newblock \emph{\bibinfo{journal}{Nature}} \textbf{\bibinfo{volume}{393}},
  \bibinfo{pages}{133--137} (\bibinfo{year}{1998}).

\bibitem{Kettle:2004hn}
\bibinfo{author}{Kettle, L.~M.}, \bibinfo{author}{Goan, H.~S.},
  \bibinfo{author}{Smith, S.~C.}, \bibinfo{author}{Hollenberg, L. C.~L.} \&
  \bibinfo{author}{Wellard, C.~J.}
\newblock \bibinfo{title}{{The effects of J-gate potential and interfaces on
  donor exchange coupling in the Kane quantum computer architecture}}.
\newblock \emph{\bibinfo{journal}{Journal of Physics: Condensed Matter}}
  \textbf{\bibinfo{volume}{16}}, \bibinfo{pages}{1011--1023}
  (\bibinfo{year}{2004}).

\bibitem{Koiller:2004gs}
\bibinfo{author}{Koiller, B.}, \bibinfo{author}{Capaz, R.},
  \bibinfo{author}{Hu, X.} \& \bibinfo{author}{Das~Sarma, S.}
\newblock \bibinfo{title}{{Shallow-donor wave functions and donor-pair exchange
  in silicon: Ab initio theory and floating-phase Heitler-London approach}}.
\newblock \emph{\bibinfo{journal}{Physical Review B}}
  \textbf{\bibinfo{volume}{70}}, \bibinfo{pages}{115207}
  (\bibinfo{year}{2004}).

\bibitem{Koiller:2001gw}
\bibinfo{author}{Koiller, B.}, \bibinfo{author}{Hu, X.} \&
  \bibinfo{author}{Das~Sarma, S.}
\newblock \bibinfo{title}{{Exchange in Silicon-Based Quantum Computer
  Architecture}}.
\newblock \emph{\bibinfo{journal}{Physical Review Letters}}
  \textbf{\bibinfo{volume}{88}}, \bibinfo{pages}{027903}
  (\bibinfo{year}{2001}).

\bibitem{Wellard:2004fb}
\bibinfo{author}{Wellard, C.~J.}, \bibinfo{author}{Hollenberg, L. C.~L.},
  \bibinfo{author}{Kettle, L.~M.} \& \bibinfo{author}{Goan, H.~S.}
\newblock \bibinfo{title}{{Voltage control of exchange coupling in phosphorus
  doped silicon}}.
\newblock \emph{\bibinfo{journal}{Journal of Physics: Condensed Matter}}
  \textbf{\bibinfo{volume}{16}}, \bibinfo{pages}{5697--5704}
  (\bibinfo{year}{2004}).

\bibitem{Wellard:2003jq}
\bibinfo{author}{Wellard, C.} \emph{et~al.}
\newblock \bibinfo{title}{{Electron exchange coupling for single-donor
  solid-state spin qubits}}.
\newblock \emph{\bibinfo{journal}{Physical Review B}}
  \textbf{\bibinfo{volume}{68}}, \bibinfo{pages}{195209}
  (\bibinfo{year}{2003}).

\bibitem{Smith:2015fo}
\bibinfo{author}{Smith, J.~S.} \emph{et~al.}
\newblock \bibinfo{title}{{Electronic transport in Si:P $\delta$-doped wires}}.
\newblock \emph{\bibinfo{journal}{Physical Review B}}
  \textbf{\bibinfo{volume}{92}}, \bibinfo{pages}{235420}
  (\bibinfo{year}{2015}).

\bibitem{Drumm:2013bs}
\bibinfo{author}{Drumm, D.~W.} \emph{et~al.}
\newblock \bibinfo{title}{{Ab~InitioElectronic Properties of Monolayer
  Phosphorus Nanowires in Silicon}}.
\newblock \emph{\bibinfo{journal}{Physical Review Letters}}
  \textbf{\bibinfo{volume}{110}}, \bibinfo{pages}{126802}
  (\bibinfo{year}{2013}).

\bibitem{Greentree:2004di}
\bibinfo{author}{Greentree, A.~D.}, \bibinfo{author}{Cole, J.~H.},
  \bibinfo{author}{Hamilton, A.~R.} \& \bibinfo{author}{Hollenberg, L. C.~L.}
\newblock \bibinfo{title}{{Coherent electronic transfer in quantum dot systems
  using adiabatic passage}}.
\newblock \emph{\bibinfo{journal}{Physical Review B}}
  \textbf{\bibinfo{volume}{70}}, \bibinfo{pages}{235317}
  (\bibinfo{year}{2004}).

\bibitem{Wang:2016ha}
\bibinfo{author}{Wang, Y.} \emph{et~al.}
\newblock \bibinfo{title}{{Highly tunable exchange in donor qubits in
  silicon}}.
\newblock \emph{\bibinfo{journal}{npj Quantum Information}}
  \textbf{\bibinfo{volume}{2}}, \bibinfo{pages}{16008} (\bibinfo{year}{2016}).

\bibitem{Tyryshkin:2012fi}
\bibinfo{author}{Tyryshkin, A.~M.} \emph{et~al.}
\newblock \bibinfo{title}{{Electron spin coherence exceeding seconds in
  high-purity silicon}}.
\newblock \emph{\bibinfo{journal}{Nature Materials}}
  \textbf{\bibinfo{volume}{11}}, \bibinfo{pages}{143--147}
  (\bibinfo{year}{2012}).

\bibitem{Knill:2001is}
\bibinfo{author}{Knill, E.}, \bibinfo{author}{Laflamme, R.} \&
  \bibinfo{author}{Milburn, G.~J.}
\newblock \bibinfo{title}{{A scheme for efficient quantum computation with
  linear optics}}.
\newblock \emph{\bibinfo{journal}{Nature}} \textbf{\bibinfo{volume}{409}},
  \bibinfo{pages}{46--52} (\bibinfo{year}{2001}).

\bibitem{Hong:1987gm}
\bibinfo{author}{Hong, C.~K.}, \bibinfo{author}{Ou, Z.~Y.} \&
  \bibinfo{author}{Mandel, L.}
\newblock \bibinfo{title}{{Measurement of subpicosecond time intervals between
  two photons by interference}}.
\newblock \emph{\bibinfo{journal}{Physical Review Letters}}
  \textbf{\bibinfo{volume}{59}}, \bibinfo{pages}{2044--2046}
  (\bibinfo{year}{1987}).

\bibitem{Lanyon:2008gv}
\bibinfo{author}{Lanyon, B.~P.} \emph{et~al.}
\newblock \bibinfo{title}{{Quantum computing using shortcuts through higher
  dimensions}}.
\newblock \emph{\bibinfo{journal}{Nature Physics}} \bibinfo{pages}{134--140}
  (\bibinfo{year}{2008}).
\newblock \eprint{0804.0272}.

\bibitem{Greentree:2004im}
\bibinfo{author}{Greentree, A.~D.} \emph{et~al.}
\newblock \bibinfo{title}{{Maximizing the Hilbert Space for a Finite Number of
  Distinguishable Quantum States}}.
\newblock \emph{\bibinfo{journal}{Physical Review Letters}}
  \textbf{\bibinfo{volume}{92}}, \bibinfo{pages}{097901}
  (\bibinfo{year}{2004}).

\end{thebibliography}

    \end{document}